\documentclass[twocolumn]{aastex701}

\usepackage{graphicx}
\usepackage{amsmath}
\usepackage{subcaption}

\providecommand{\sw}[1]{\texttt{#1}}
\newcommand{\jetsimpy}{\sw{jetsimpy}}

\newcommand{\threeml}{\sw{3ML}}

\newcommand{\git}{GROWTH-India Telescope}
\newcommand{\ztf}{Zwicky Transient Facility}
\newcommand{\hct}{Himalayan Chandra Telescope}
\newcommand{\ugmrt}{uGMRT}
\newcommand{\jvla}{JVLA}

\newcommand{\swift}{\emph{Swift}}
\newcommand{\fermi}{\emph{Fermi}}

\newcommand{\astrosat}{\emph{AstroSat}}

\newcommand{\chandra}{\emph{Chandra}}

\newcommand{\thisgrb}{GRB\,230812B}
\newcommand{\thissn}{SN\,2023pel}
\newcommand{\tnt}{T\(_{90}\)}
\newcommand{\pf}{\ensuremath{\Pi}}
\newcommand{\Gzero}{\ensuremath{\Gamma_0}}
\newcommand{\thetaj}{\(\theta_j\)}
\newcommand{\thetav}{\(\theta_v\)}

\newcommand{\ekiso}{\(E_{k,{\rm iso}}\)}

\newcommand{\rband}{r\(^\prime\)}
\newcommand{\gband}{g\(^\prime\)}
\newcommand{\iband}{i\(^\prime\)}
\newcommand{\zband}{z\(^\prime\)}
\newcommand{\uband}{u\(^\prime\)}

\begin{document}
\title{Can We Find the Emission Mechanism Behind the Extremely Bright \thisgrb?}

\author[0009-0002-7897-6110]{Utkarsh Pathak}
\email{utkarshpathak.07@gmail.com}
\affiliation{Department of Physics, Indian Institute of Technology Bombay, Powai, Mumbai 400076, India}

\author[0009-0002-8110-0515]{Sameer K. Patil}
\email{sameerkpatil@gmail.com}
\affiliation{Department of Physics, Indian Institute of Technology Bombay, Powai, Mumbai 400076, India}

\author[0009-0008-6644-5412]{Hitesh Tanenia}
\email{hitesh.tanenia@iitb.ac.in}
\affiliation{Department of Physics, Indian Institute of Technology Bombay, Powai, Mumbai 400076, India}

\author[0009-0005-5080-0107]{Yashowardhan Rai}
\email{yash_rai@iitb.ac.in}
\affiliation{Department of Physics, Indian Institute of Technology Bombay, Powai, Mumbai 400076, India}

\author[0000-0002-7942-8477]{Viswajeet Swain}
\email{vishwajeet.s@iitb.ac.in}
\affiliation{Department of Physics, Indian Institute of Technology Bombay, Powai, Mumbai 400076, India}

\author[0009-0007-9244-191X]{Anuraag Arya}
\email{arya.a@iitb.ac.in}
\affiliation{Department of Physics, Indian Institute of Technology Bombay, Powai, Mumbai 400076, India}

\author[0000-0003-3630-9440]{Gaurav Waratkar}
\email{gauravw@caltech.edu}
\affiliation{Department of Physics, Indian Institute of Technology Bombay, Powai, Mumbai 400076, India}

\author[0000-0003-3173-4691]{Anirudh Salgundi}
\email{anirudhs@unc.edu}
\affiliation{Department of Physics and Astronomy, University of North Carolina at Chapel Hill, 120 E. Cameron Ave., Chapel Hill, NC 27514, USA}
\affiliation{Department of Physics, Indian Institute of Technology Bombay, Powai, Mumbai 400076, India}

\author[0009-0004-9984-4138]{Mehul Goyal}
\email{mehul.goyal@iitb.ac.in}
\affiliation{Department of Physics, Indian Institute of Technology Bombay, Powai, Mumbai 400076, India}

\author[0000-0002-6112-7609]{Varun Bhalerao}
\email{varunb@iitb.ac.in}
\affiliation{Department of Physics, Indian Institute of Technology Bombay, Powai, Mumbai 400076, India}

\author[0000-0002-8977-1498]{Igor Andreoni}
\email{igor.andreoni@unc.edu}
\affiliation{Department of Physics and Astronomy, University of North Carolina at Chapel Hill, 120 E. Cameron Ave., Chapel Hill, NC 27514, USA}

\author[0000-0002-7686-3334]{Sarah Antier}
\email{antier@ijcab.in2p3.fr}
\affiliation{IJCLab, Univ Paris-Saclay, CNRS/IN2P3, Orsay, France}

\author[0000-0003-3533-7183]{G. C. Anupama}
\email{gca@iiap.res.in}
\affiliation{Indian Institute of Astrophysics, II Block Koramangala, Bengaluru 560034}

\author[0000-0002-3927-5402]{Sudhanshu Barway}
\email{sudhanshu.barway@iiap.res.in}
\affiliation{Indian Institute of Astrophysics, II Block Koramangala, Bengaluru 560034}

\author[0000-0003-3352-3142]{Dipankar Bhattacharya}
\email{dipankar.bhattacharya@ashoka.edu.in}
\affiliation{Department of Physics, Ashoka University, Sonepat, Haryana 131029, India}

\author[0000-0001-9856-1866]{Tanmoy Chattopadhyay}
\email{tanmoyc@stanford.edu}
\affiliation{Kavli Institute for Particle Astrophysics and Cosmology, Stanford University, 452 Lomita Mall, Stanford, CA 94305, USA}

\author[0000-0002-8262-2924]{Michael Coughlin}
\email{cough052@umn.edu}
\affiliation{School of Physics and Astronomy, University of Minnesota, Minneapolis, Minnesota 55455, USA}

\author[0000-0002-7708-3831]{Anirban Dutta}
\email{anirbandutta@astro.ncu.edu.tw}
\affiliation{Graduate Institute of Astronomy, National Central University, 300 Zhongda Road, 32001 Zhongli, Taiwan}

\author[0000-0002-4223-103X]{Christoffer Fremling}
\email{fremling@caltech.edu}
\affiliation{Caltech Optical Observatories, California Institute of Technology, Pasadena, CA 91125, USA}
\affiliation{Division of Physics, Mathematics and Astronomy, California Institute of Technology, Pasadena, CA 91125, USA}

\author[0000-0003-1585-8205]{Nidhal Guessoum}
\email{nguessoum@aus.edu}
\affiliation{Physics Department, American University of Sharjah, Sharjah, UAE}

\author[0000-0002-5619-4938]{Mansi Kasliwal}
\email{mansi@astro.caltech.edu}
\affiliation{Division of Physics, Mathematics and Astronomy, California Institute of Technology, Pasadena, CA 91125, USA}

\author[0000-0002-7150-9061]{Oliver J. Roberts}
\email{oliver.roberts@universityofgalway.ie}
\affiliation{Science and Technology Institute, Universities Space and Research Association, 320 Sparkman Drive, Huntsville, AL 35805, USA}
\altaffiliation{Physics, School of Natural Sciences, University Road, University of Galway, Galway, Ireland, H91 TK33}

\author[0000-0002-6428-2700]{Gokul P. Srinivasaragavan}
\email{gsriniv2@umd.edu}
\affiliation{Department of Astronomy, University of Maryland, College Park, MD 20742, USA}

\author[0000-0002-0525-0872]{Rishabh Singh Teja}
\email{rsteja@sjtu.edu.cn}
\affiliation{Tsung-Dao Lee Institute, Shanghai Jiao Tong University, 1 Lisuo Road, Shanghai 201210, China}

\author[0000-0002-2050-0913]{Santosh Vadawale}
\email{santoshv@prl.res.in}
\affiliation{Physical Research Laboratory, Ahmedabad, Gujarat 380009, India}

\author[0000-0002-2149-9846]{Peter Veres}
\email{pv0004@uah.edu}
\affiliation{Department of Space Science, University of Alabama in Huntsville, Huntsville, AL 35899, USA}
\affiliation{Center for Space Plasma and Aeronomic Research, University of Alabama in Huntsville, Huntsville, AL 35805, USA}


\begin{abstract}
\thisgrb\ is a bright long-duration GRB with a luminous, long-lived afterglow and an \astrosat/CZTI polarization measurement during the prompt phase, enabling a joint study of its prompt spectral evolution, polarization, and broadband afterglow. Time-resolved spectroscopy of the prompt emission shows that during the rising phase, the low-energy Band-function index exceeds the synchrotron line of death, favoring the presence of an additional thermal component. At later times, from $T_0+2$~s to $T_0+32$~s, the prompt spectra are consistent with predominantly non-thermal emission. Polarization analysis of the prompt emission in the $300$--$600$~keV band yields a marginal lower limit on the polarization fraction of $\Pi \gtrsim 50\%$ at the $1\sigma$ level.

The long X-ray monitoring of the afterglow shows no jet break over the observed baseline. Multiwavelength afterglow modeling favors a wide jet with an inferred half-opening angle of \(\theta_j = 15^{+6}_{-4}\)~degrees observed close to the jet axis with a viewing angle of \(\theta_v = 0.9^{+1.8}_{-0.6}\)~degrees. The inferred circumburst density is low, \(n_0 = 1.2^{+0.3}_{-0.1}\times10^{-4}\,\textrm{cm}^{-3}\), and the isotropic-equivalent kinetic energy of the jet is \ekiso\(= 4.0^{+1.5}_{-0.8} \times 10^{53}\)~erg. 

Taken together, the prompt spectral evolution favors an early phase with a thermal contribution followed by a later phase dominated by non-thermal emission. The polarization constraint in the late prompt phase is consistent with synchrotron emission, although a higher-significance polarization measurement will be required to robustly constrain the magnetic-field geometry and the relative contribution of photospheric emission.
\end{abstract}


\section{Introduction}\label{sec:introduction}
Gamma-ray bursts (GRBs) are extremely bright transients in the universe, typically lasting from fractions of a second to several minutes. One conventionally used classification separates GRBs based on the duration into short GRBs (\tnt\ \(\lesssim 2\)~s) and long GRBs (\tnt\ \(\gtrsim 2\)~s;~\citealt{Kouveliotou1993, KumarZhang2015}). Another physically motivated classification distinguishes compact-object merger-driven GRBs (type I), with harder spectra, from core-collapse-driven GRBs (type II), which are commonly associated with softer spectra. Although these classifications broadly overlap, there are a few exceptions, including short-duration collapsar~\citep[e.g.,][]{short_collapsar} and long-duration merger events such as GRB\,230307A and GRB\,211121A~\citep{230307A, 21112A}.

Several open questions concern the nature of the central engine, the mechanism that launches relativistic jets. The presence and topology of magnetic fields within the relativistic outflow strongly influence jet acceleration, dissipation, and radiation processes~\citep{Lyutikov2003,ZhangYan2011,KumarZhang2015}. Various emission-mechanism models can describe the prompt emission seen in GRBs; these models are often spectrally degenerate. The non-thermal radiation observed during the prompt phase may arise from synchrotron or inverse Compton emission, whereas the photospheric model can describe the thermal radiation seen in some GRBs.

A promising way to break the spectral degeneracy among these models is to measure the high-energy polarization of the prompt phase~\citep{toma2009,lundman2014,gill2021,CovinoGotz2016}. The fraction of light being linearly polarized (polarization fraction, hereafter \pf) is determined by the intrinsic emission mechanism and magnetic field structures. Theoretical studies predict that different emission mechanisms, together with their magnetic field configurations, such as Synchrotron with Ordered magnetic field (SO), Synchrotron with Randomly oriented magnetic field (SR), Compton Drag (CD), and photospheric emission (P), produce different polarization signatures~\citep{gill2021,lundman2014,toma2009}. Measurements on \pf\ within a given energy band can be used to favor or disfavor certain models, thereby providing a powerful, independent constraint that complements spectral and temporal analyses. Detecting polarization is extremely difficult, even upper or lower limits on the polarization fraction can exclude portions of parameter space for several prompt-emission scenarios~\citep{toma2009,gill2021,sharma2020,2019NatAs...3..258Z}.

The observed \pf\ depends on the composition, magnetization, and geometry of the jet. Jet interacts with the ISM and decelerates, producing broadband afterglow emission~\citep{MeszarosRees1997, Sari1998}. The broadband afterglow evolution, in particular the presence or absence of a jet break, carries information regarding the jet, viewing geometry, circumburst medium, kinetic energy of the outflow \ekiso, and the initial bulk Lorentz factor \Gzero~\citep{2018ApJ...859..160W,2010ApJ...725.2209L}. The macrophysical and microphysical properties, such as jet half‑opening angle \thetaj, viewing angle \thetav, and initial bulk Lorentz factor \(\Gamma_0\), can be inferred from afterglow modeling. 
When the jet is viewed close to its edge or moderately off-axis, geometric asymmetry can produce a relatively high polarization fraction even for photospheric emission models~\citep{lundman2014}. In contrast, for an on-axis observer viewing a wide emitting region (\(\Gamma^{-1} \sim \theta_j\)), polarization vectors from different parts of the jet can partially cancel, leading to a low net \pf, even in the presence of globally ordered magnetic fields. When the relativistic beaming cone is much narrower than the jet opening angle (\(\Gamma^{-1} \ll \theta_j\)), the observer probes only a localized region of the jet, reducing geometric cancellation and allowing a higher observed \pf\ depending on the underlying magnetic-field structure~\citep{toma2009,gill2021}. Thus, the observed polarization \pf\ strongly depends on the jet geometry and viewing configuration, motivating extensive broadband afterglow follow-up for GRBs with prompt polarization measurements.

\astrosat/CZTI has enabled several polarization studies in the hard X-ray regime~\citep{chattopadhyay2019,sharma2020,czti_nature}, which disentangle spectrally degenerate emission mechanisms. However, many of them don't have extensive afterglow data. Such joint studies remain limited because only a small fraction of GRBs are bright enough for high-energy polarization and have an extensive broadband afterglow coverage for jet modeling. 

\thisgrb\ presents a rare opportunity where both conditions are satisfied, with prompt polarization constraints from \astrosat/CZTI and late-time broadband afterglow observations from \chandra. \thisgrb\ is associated with the supernova \thissn. A detailed analysis of the supernova properties and the GRB--SN connection is presented in the companion work by~\citet{sn2023pel}, while the present work focuses on the prompt polarization properties and broadband afterglow modeling of the jet.

This paper is structured as follows: Section~\ref{sec:observations} summarizes the observations and data collection for \thisgrb, presented sequentially for the prompt and afterglow phases. Section~\ref{sec:prompt} details the prompt spectral and polarization analysis. Section~\ref{sec:afterglow-analysis} presents the broadband afterglow data, a simple power-law fitting to the data, and detailed afterglow modeling. Section~\ref{sec:discussion} discusses the polarization detection, inferred jet geometry, and their implications on plausible emission mechanisms.

\begin{figure*}[hbt!]
    \centering
    \includegraphics[width = \linewidth, height = 0.333\linewidth]{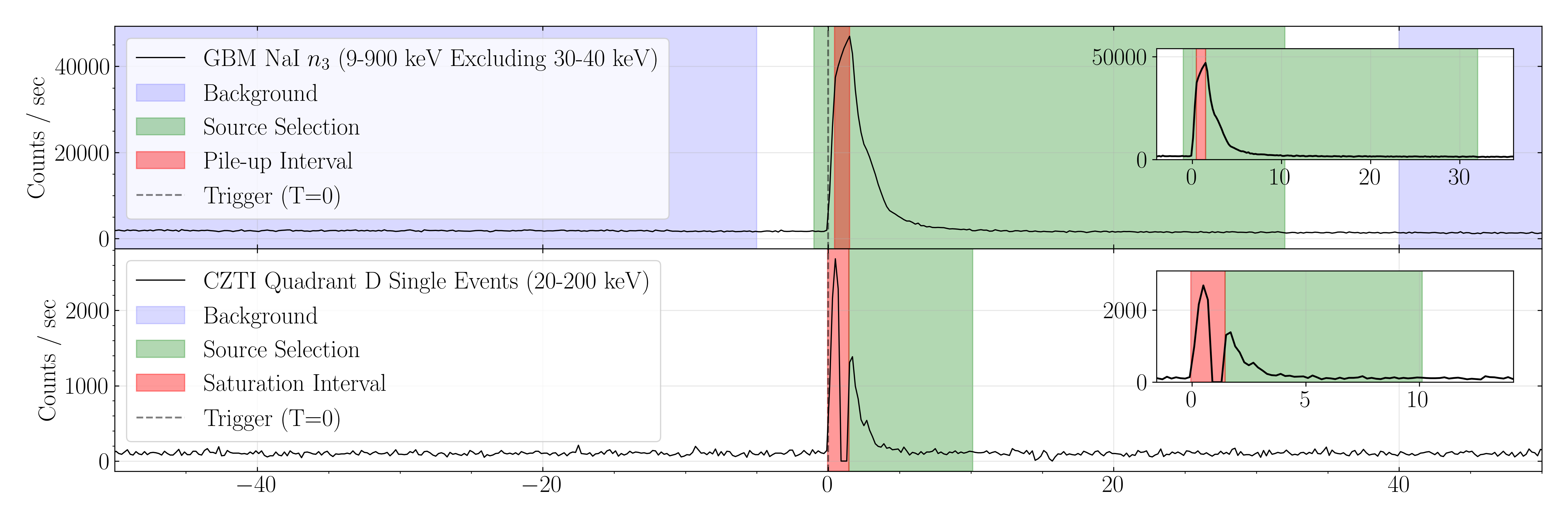}
    \caption{The prompt light curves from the brightest detector (\(n_3\)) in \fermi/GBM and quadrant D of \astrosat/CZTI. The \(30\)--\(40\)~keV energy range is excluded because of the iodine K-edge in \fermi/GBM. The red-shaded region indicates the saturated portion of the burst, which differs between the two instruments. The green-shaded region marks the source interval adopted for the respective \(T_{90}\) analyses of \fermi/GBM~\citep{gcn34391} and \astrosat/CZTI. The background intervals are shown in blue for \fermi/GBM; for \astrosat/CZTI, the background is estimated from \(-300\)~s to \(-70\)~s before the trigger and from \(70\)~s to \(300\)~s after the trigger. See Section~\ref{sec:pol} for further details on saturation in \astrosat/CZTI quadrants.\label{fig:promptlc}}
\end{figure*}

\section{Observations and Data}\label{sec:observations}
\subsection{Fermi}
\thisgrb\ was first detected by \fermi/GBM~\citep{fermigbm} at 18:58:12 UTC on 2023 August 12~\citep[hereafter \(T_0\);][]{gcn34386}. The localization is \({\rm RA} = 250.06^\circ\), \({\rm DEC} = 46.2^\circ\) with \(1^\circ\) error radius, with a fluence of \(2.52 \times 10^{-4}~\mathrm{erg~cm}^{-2}\) in \(10\)--\(1000\)~keV~\citep{gcn34391}. The GBM light curve shows a narrow peak (\tnt\ $= 3$~s) with emission extending to ${\sim}20$~s~\citep[Figure~\ref{fig:promptlc};][]{gcn34386,gcn34387,gcn34391}. \fermi/LAT~\citep{fermilat} detected spatially correlated high-energy emission up to 72~GeV at ${\sim}32.2$~s post-trigger, with a boresight angle of $29^\circ$ at trigger time~\citep{gcn34392}.

 \subsection{AstroSat}\label{astrosat-obs}
Cadmium Zinc Telluride Imager, on-board \astrosat~\citep[\astrosat/CZTI;][]{CZTIvb}, detected the GRB at an angle of \(\theta \approx 23^\circ\) and \(\phi \approx 121^\circ\) in CZTI frame, within the allowed range of \(\theta\ \leq 60 ^\circ\) for GRB to be used in polarization analysis~\citep{AstroSatpolarization_5yrcatalog}. Polarization analysis is discussed in Section~\ref{sec:pol}. The bright emission saturated the detectors (Figure~\ref{fig:promptlc}). However, the unsaturated tail of the GRB was detected with \tnt\ of tail \(\approx 9\)~s in the energy range \(20\)--\(200\)~keV (Figure~\ref{fig:czti_quad_saturate}). \astrosat/LAXPC also detected the burst with \tnt\ \(\approx 4\)~s for energy range \(30\)--\(400\)~keV \citep{gcn34526}.
 
 
\subsection{Swift}
\swift/XRT~\citep{burrows2005} localized the afterglow at RA (J2000): $16^h 36^m 31.76^s$ and Dec (J2000): $47\degr 51\arcmin 26.7\arcsec$ with an error radius of 3.7~arcsec~\citep{gcn34394}. The observation started at $T_0 + 25$~ks, continuing through the end of August 2023 with ${\sim}5$~ks exposures in the $0.3$--$10.0$~keV band~\citep{gcn34393,gcn34400}. The XRT data were processed using the automated analysis pipeline of the UK Swift Science Data Centre to produce light curves and spectra~\citep{evans2009}. \swift/UVOT detected the afterglow in the white filter with a total exposure of $4.8$~ks in the first few epochs~\citep{gcn34399}.

\subsection{Chandra}
We triggered \chandra\ for late X-ray afterglow observations through Chandra DDT~\citep[proposal number \(24408929\), PI Pathak;][]{gcn34632}. The source was observed for a total of \(50.17\)~ks effective exposure in three epochs from \(3~\textrm{September}~2023\) to \(22~\textrm{September}~2023\), detecting the afterglow till \(41.44\)~days from trigger time. \chandra\ consists of an advanced CCD imaging spectrometer where we have selected \(4\) ACIS-S cameras for our observation: a configuration for higher sensitivity but a smaller Field of View (FoV) suited for an already well-localized GRB. The images, spectra, and light curves were made from the \chandra\ level-2 data, reprocessed using the \texttt{CIAO 4.15} module. This research employs a collection of datasets obtained with the Chandra X-ray Observatory, available in \dataset[DOI: 10.25574/cdc.636]{https://doi.org/10.25574/cdc.636}.

\subsection{Optical Afterglow Follow-up}
The optical afterglow was identified as ZTF23aaxeacr by the Zwicky Transient Facility~\citep[ZTF;][]{2019PASP..131a8002B,2019PASP..131g8001G,2020PASP..132c8001D,2019PASP..131a8003M}, with the first detection ${\sim}8.6$~hr post GBM-trigger in \rband\ and \gband; the associated supernova \thissn\ was subsequently reported to Transient
Name Server (TNS)~\citep{sn2023pel}. We use the ZTF Fritz marshal to plan follow-up observations and to store photometric data~\citep{2019JOSS....4.1247V, Coughlin_2023}. The GROWTH-India Telescope~\citep[GIT;][]{kumar2022} began observing within ${\sim}20$~hr of the GBM-trigger~\citep{gcn34420} in \rband, \gband, and \iband, with $300$~s exposures continuing for 7~days; stacked images were used to reach fainter magnitudes at late times~\citep[for detailed analysis procedures see][]{pathak2026}. Additional photometric follow-up was obtained with the 2~m Himalayan Chandra Telescope (HCT) in \rband\ and \iband\ from 3 to 12 days post GBM-trigger, the 4.3~m Lowell Discovery Telescope (LDT) with the Large Monolithic Imager (LMI) in \rband\ and \iband\ over 5 epochs in late August to September 2023, and the 2~m robotic Liverpool Telescope~\citep[LT;][]{Steele2004} in \gband, \rband, \iband, and \zband\ over three nights. Photometry from the \textsc{P60} observations was obtained using the image-subtraction pipeline described in~\citet{fremling2016}. For all optical telescopes, images were corrected for bias, flat field, and cosmic rays using \sw{Astro-SCRAPPY}~\citep{2019ascl.soft07032M}, astrometrically solved with \sw{solve-field}~\citep{2010AJ....139.1782L}, and stacked with \sw{SWARP} where needed. Host contributions were removed via image subtraction using the \sw{ZOGY} algorithm, and PSF-fitted photometry was calibrated against PS1~DR1 standard stars~\citep{swain2025,sn2023pel}. A comprehensive description of the optical data acquisition and subsequent reduction is presented in our companion paper~\citep{sn2023pel}. Furthermore, to construct a robust dataset for our afterglow modeling, we incorporate supplementary optical photometry previously published by~\citet{grandma}.
 
\subsection{Redshift}
\citet{gcn34409} measured a spectroscopic redshift of $z = 0.36$ from absorption (Mg\,{\sc ii}, Mg\,{\sc i}, Ca\,{\sc ii}, Ca\,{\sc i}) and emission
([O\,{\sc ii}], [O\,{\sc iii}]) lines using OSIRIS+ on the 10.4~m GTC. Throughout this work we adopt \sw{Planck2018} cosmology~\citep{planck2018} with
$H_0 = 67.66~\mathrm{km~s^{-1}~Mpc^{-1}}$, $\Omega_m = 0.310$, $\Omega_\Lambda = 0.689$, giving a luminosity distance of $1.98$~Gpc.
 
\subsection{Radio Afterglow Follow-up}
We obtained observations using the Karl G. Jansky Very Large Array~\citep[\jvla;][]{gcn34735} and the upgraded Giant Metrewave Radio Telescope~\citep[\ugmrt;][]{gcn34735}. To construct a comprehensive broadband dataset, we supplemented our observations with other radio detections, including earlier \jvla\ data~\citep{gcn34552}, Northern Extended Millimeter Array~\citep[NOEMA;][]{gcn34468}, and Arcminute Microkelvin Imager Large Array~\citep[AMI-LA;][]{gcn34433} observations. The combined radio dataset spans $1.26$--$90$~GHz from 2 to 35.7~days post GBM-trigger. We note that our observation~\citep{gcn34735} has a detection in 6~GHz at $\sim$21~days, while ~\citet{gcn35505} has found an upper limit at 17~days in 6~GHz.

\section{Prompt emission: analysis}\label{sec:prompt}
We present the analysis of the spectra and polarization of the prompt phase in Section~\ref {sec:prompt_spec} and Section~\ref {sec:pol}, respectively. 

\subsection{Spectral Analysis}\label{sec:prompt_spec}
The prompt emission of \thisgrb\ has been analyzed in two independent studies. \citet{ror2024promptafterglowanalysisfermilat} find that a Band + blackbody model~\citep{band1993} best describes the time-integrated spectrum, with a blackbody temperature $kT \approx 24$~keV superimposed on the non-thermal continuum, which they interpret as a photospheric contribution during the early, bright phase. \citet{mohnani2026broad} model the keV--GeV spectral energy distribution with Band + synchrotron self Compton (SSC), attributing the 72~GeV \fermi/LAT photon detected at $T_0 + 32$~s to an SSC or inverse Compton (IC) component peaking in the GeV band~\citep{gcn34392}.

\begin{figure}[hbt!]
    \centering
    \includegraphics[width=0.9\linewidth]{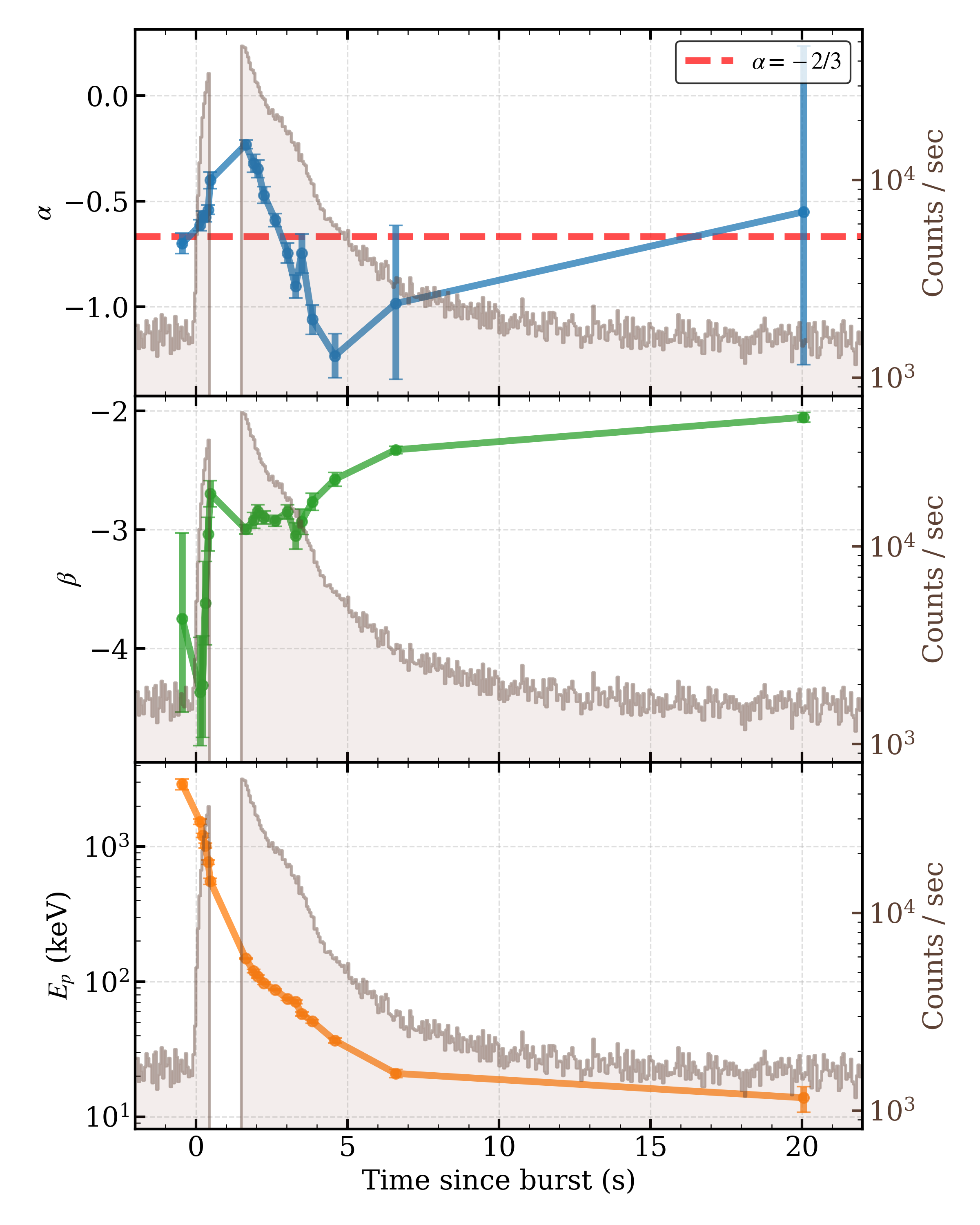}
    \caption{Spectral parameter evolution across the burst using Band function. \(E_P\) shows a monotonic decrease across the burst. \(\alpha\) during rise phase, crosses the synchrotron line-of-death at \(-2/3\) shown as red dashed line. Background light curve is plotted for the \fermi/GBM detector \(n_3\). \label{fig:TR_spec_evolution}}
\end{figure}

We applied the Bayesian block algorithm to the brightest detector ($n_3$) to define time-resolved spectral intervals. Adjacent bins with signal-to-noise ratio $<50$ were merged, yielding 18 time bins (Figure~\ref{fig:TR_spec_evolution}). We selected 4 NaI detectors: $n_0$, $n_3$, $n_6$, and $n_7$ based on their viewing angles. The remaining detectors were excluded as their viewing angles exceeded $60\degr$. To extend the spectral coverage to higher energies, we also included the BGO detector $b_0$. The data were fit with the Band function over the energy range $9$--$900$~keV, excluding the $30$--$40$~keV region (corresponding to the iodine K-edge) for the NaI detectors and $250$--$38000$~keV for the BGO detector, within the \threeml~\citep{2015arXiv150708343V} framework.
 
In our study, we find the prompt spectrum to evolve strongly with time: the early phase ($T_0 - 1$ to $T_0 + 2$~s) is characterized by $\alpha \le -0.25$, well above the synchrotron line-of-death at $-2/3$~\citep{preece1998synchrotron}, consistent with a subdominant thermal contribution to the non-thermal continuum~\citep{ryde2005}. By contrast, the late-time emission ($T_0 + 2$~s to $T_0 + 32$~s) is well described by the Band function alone, with spectral indices typical of optically thin synchrotron emission in long GRBs~\citep{fermi10yr}.

The spectral evolution is shown in Figure~\ref{fig:TR_spec_evolution}. $E_p$ decreases monotonically across the burst (hard-to-soft evolution), indicative of adiabatic cooling of the emission region, consistent with prior results~\citep{ror2024promptafterglowanalysisfermilat}. During the burst rise ($T_0 - 1$~s to $T_0 + 2$~s), $\alpha$ exceeds $-2/3$, crossing the synchrotron line-of-death~\citep{preece1998synchrotron}; this hardening is most plausibly attributed to a subdominant photospheric or thermal component~\citep{ryde2005,ror2024promptafterglowanalysisfermilat}. Alternatively, hard low-energy indices (\(\alpha > -2/3\)) can also arise within the synchrotron framework through marginally fast cooling~\citep{daigne2011,beniamini2013}, decaying magnetic fields~\citep{peer2006,uhm2014}, jitter radiation~\citep{medvedev2000,burgess2014jitter}, or low-energy spectral breaks~\citep{oganesyan2019}. Beyond $T_0 + 2$~s, $\alpha$ softens below $-2/3$, and the spectrum is well described by synchrotron alone; this is the regime used for the polarization analysis (Section~\ref{sec:pol}).

Since our primary focus is the polarimetric analysis of the CZTI unsaturated post-peak interval (Section~\ref{sec:pol}), we modeled spectra for this specific time selection. The late-time phase is free of the thermal excess and corresponds to the Band-only regime; we therefore adopt the Band function as the best-fit model. Fitting the spectrum over the polarization time interval ($T_0 + 1.5$~s to $T_0 + 10.2$~s; Section~\ref{sec:pol}), using the same NaI and BGO detectors and energy ranges, we obtain $\alpha = -0.90 \pm 0.01$, $\beta = -2.85 \pm 0.03$, and $E_p = 106 \pm 1$~keV ($E_{p,z} = 144 \pm 1$~keV).

\subsection{Polarization Analysis}\label{sec:pol}
\begin{figure}[!htb]
    \centering
    \includegraphics[width=\linewidth]{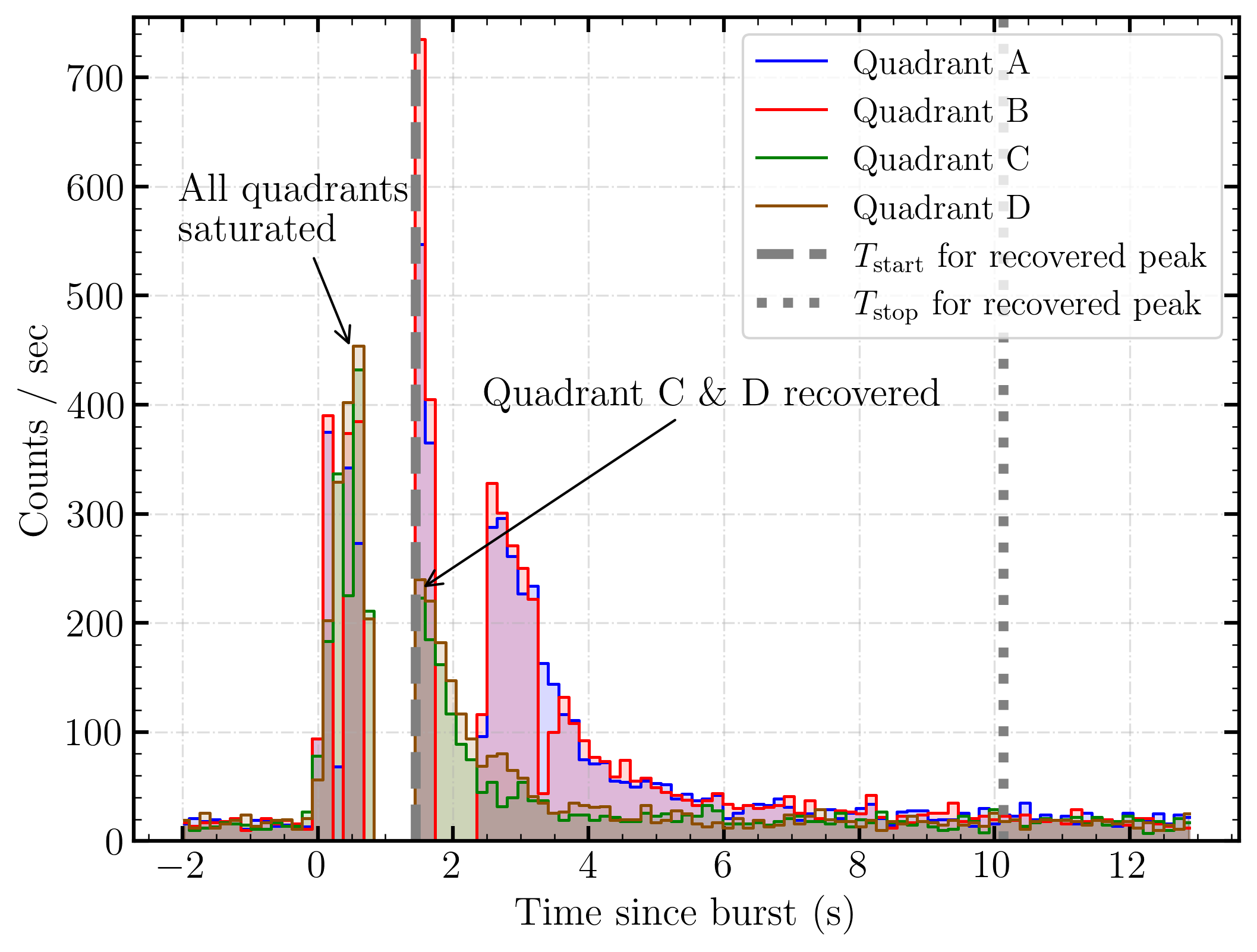}
    \caption{Quadrant-wise prompt light curve of \thisgrb\ from \astrosat/CZTI. All four quadrants are saturated during the initial rise phase. Quadrants C and D recover first at $T_0 + 1.5$~s (dashed line; $T_\mathrm{start}$), while Quadrants A and B, being source-facing, recover ${\sim}1$~s later. The dotted line at $T_\mathrm{stop} = T_0 + 10.2$~s marks the end of the source interval used for polarization analysis.\label{fig:czti_quad_saturate}}
\end{figure}
\begin{figure}[!hbt]
    \centering
    \includegraphics[width=\linewidth]{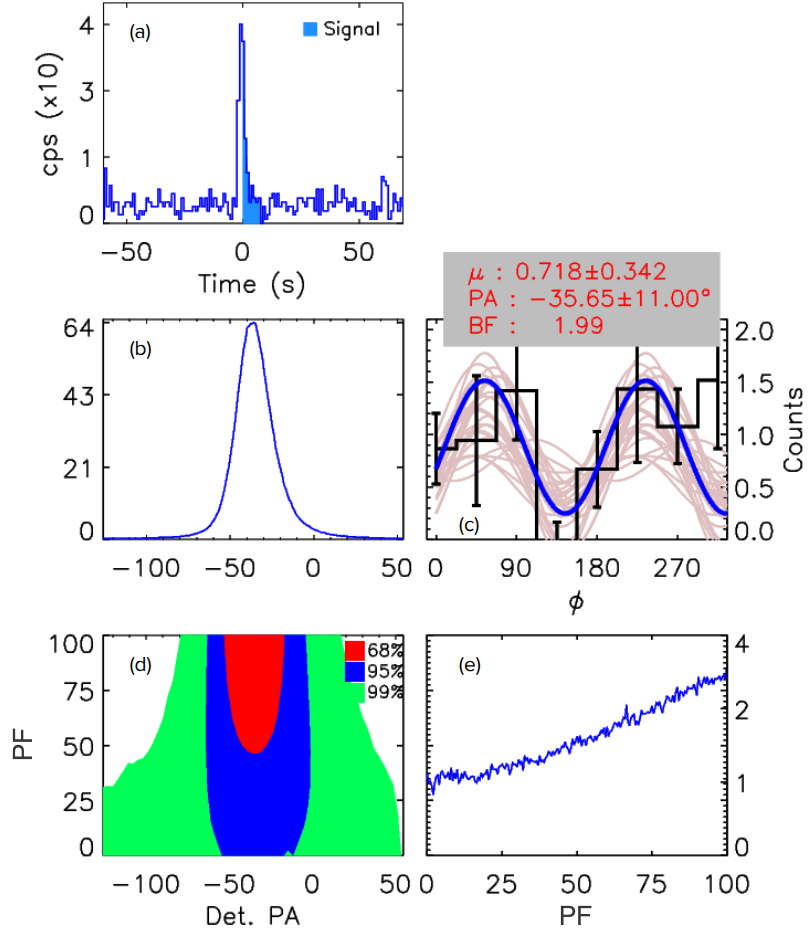}
    \caption{Polarization analysis of \thisgrb\ in the $300$--$600$~keV band with \astrosat/CZTI. (a) CZTI light curve with the signal interval (shaded) used for analysis. (b) Marginal posterior of the polarization angle (PA). (c) Azimuthal Compton scattering angle ($\phi$) distribution (black histogram) with MCMC posterior draws (pink) and the best-fit sinusoid (blue); the inferred modulation factor $\mu = 0.7\pm0.3$, $\mathrm{PA} = -36^\circ \pm 11^\circ$, and Bayes factor $\mathrm{BF} = 1.99$ are annotated. (d) Joint 68\%, 95\%, and 99\% credible contours in the PA--PF plane. (e) Marginal posterior of the polarization fraction (PF), consistent with a lower-limit measurement of $\Pi \ge 50\%$ at $1\sigma$.\label{fig:prompt_polarization}}
\end{figure}

As discussed in Section~\ref{astrosat-obs}, \thisgrb\ was detected by \astrosat/CZTI at \((\theta, \phi) \approx (23^\circ, 121^\circ)\) in the CZTI reference frame, well within the recommended range of \(0^\circ\)--\(60^\circ\) and \(120^\circ\)--\(180^\circ\) for polarization analysis~\citep{AstroSatpolarization_5yrcatalog}. For this source direction, quadrants A \& B received the highest photon flux, whereas quadrants C \& D were partially shadowed by A \& B and therefore recorded fewer photons (Figure~\ref{fig:czti_quad_saturate}). At the burst peak, all four quadrants were saturated (\(T_0\) to \(T_0+1.5\)~s). Quadrants C \& D recovered at \(T_0+1.5\)~s. Owing to how CZTI read out, quadrants A \& B recorded photons for only \(\sim0.3\)~s after \(T_0+1.5\)~s before their buffers filled again, following which no events were recorded for approximately \(1\)~s. All four quadrants resumed nominal operation after \(T_0+2.5\)~s. We therefore use the interval \(T_0+1.5\)~s to \(T_0+10.2\)~s, corresponding to the post-peak tail of the burst and coinciding with the non-thermal-dominated spectral regime identified in Section~\ref{sec:prompt_spec}. This interval yields 765 viable Compton double-event counts for the polarization analysis.

The polarization analysis follows the standard \astrosat/CZTI pipeline, which uses Compton scattering events between adjacent detector pixels to reconstruct the azimuthal scattering-angle distribution. The modulation in this distribution is characterized via MCMC sampling and Bayesian model comparison, with the Bayes factor (BF) quantifying evidence for a polarized signal over an unpolarized one~\citep{AstroSatpolarization_5yrcatalog}. In the full $100$--$600$~keV band, no significant modulation is detected; the $99\%$ upper limit on \pf\ is $\le 64\%$.

We further perform polarization analysis in different energy ranges, motivated by theoretical predictions that different emission mechanisms produce energy-dependent polarization signatures~\citep{toma2009,gill2021,2020ApJ...892..141L}, and following the approach adopted in prior \astrosat/CZTI GRB polarization studies~\citep{2024ApJ...972..166G,2019ApJ...884..123C,sharma2020}. In the $300$--$600$~keV band, we measure a modulation factor $\mu = 0.7 \pm 0.3$ and a polarization angle $\mathrm{PA} = -36^\circ \pm 11^\circ$, with a Bayes factor $\mathrm{BF} = 1.99$ in favor of a polarized signal (Figure~\ref{fig:prompt_polarization}). The marginal posterior of PF rises monotonically (Figure~\ref{fig:prompt_polarization}e), consistent with a $1\sigma$ lower limit of $\Pi \ge 50\%$. The joint PA--PF credible contours (Figure~\ref{fig:prompt_polarization}d) confirm that the PA is well-constrained while the PF posterior is consistent with a high polarization fraction. Although the BF $\approx 2$ constitutes weak-to-moderate Bayesian evidence, the $1\sigma$ lower limit of $\Pi \ge 50\%$ hints at a physically meaningful constraint on the emission mechanism, as we discuss in Section~\ref{sec:pol_discussion}.

\section{Afterglow}\label{sec:afterglow-analysis}
\begin{figure}
    \centering
    \includegraphics[width=\linewidth]{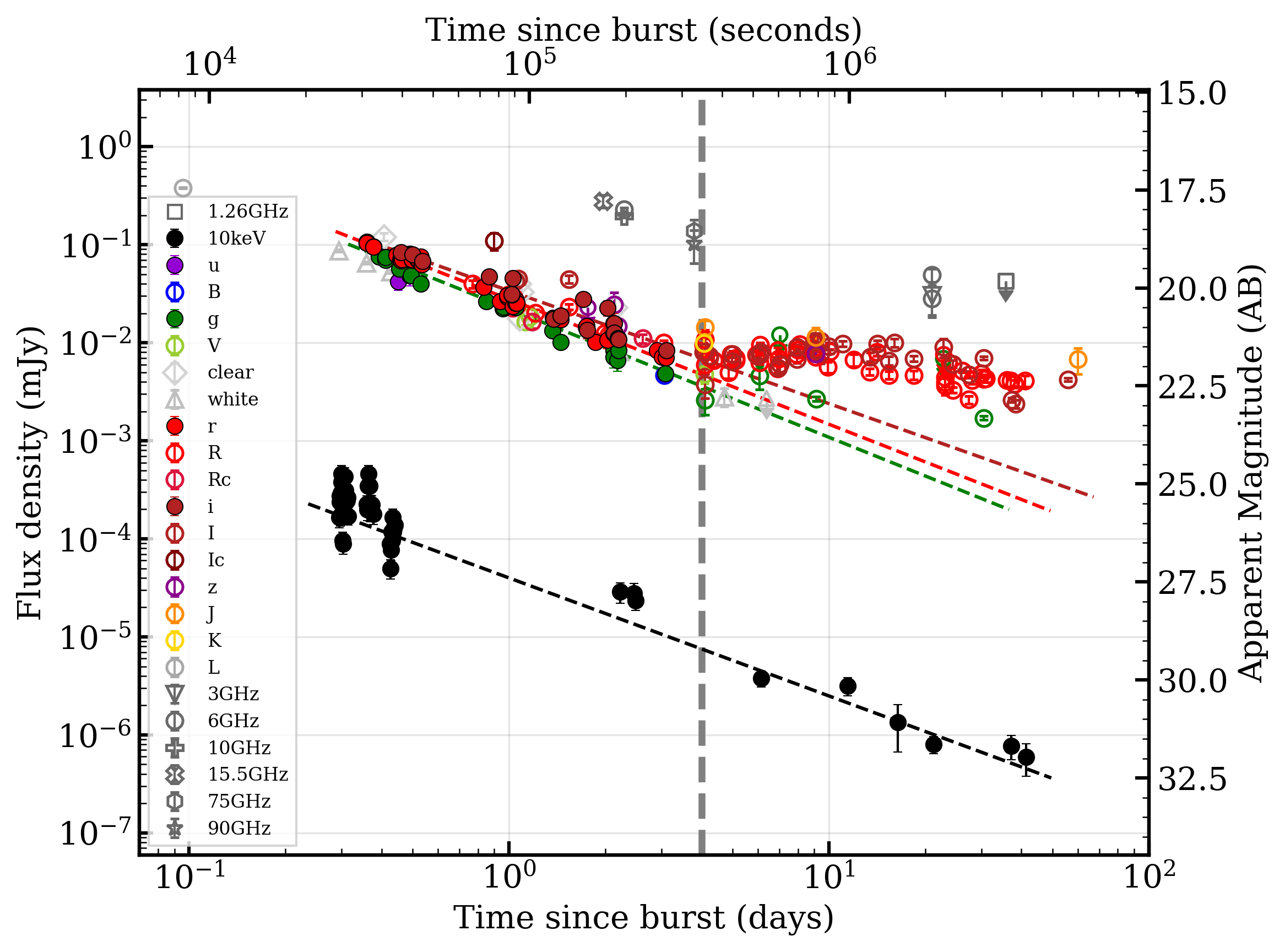}
    \caption{Multi-band afterglow light curve of \thisgrb\ in optical, radio, and X-ray regimes. Filled markers indicate data used in afterglow modeling. Supernova \thissn\ dominates in optical at $t>4$~days, shown as a vertical dashed line, and optical data beyond this line are excluded from fitting. The temporal decay indices from simple power-law fits are: $\alpha_g = 1.31 \pm 0.01$, $\alpha_r = 1.28 \pm 0.01$, $\alpha_i = 1.15 \pm 0.02$ (optical), and $\alpha_{\rm X} = 1.23 \pm 0.04$ (X-ray at 10~keV). 1.26~GHz data is shown as an upper limit.\label{fig:afterglow-simple-powerlaw}}
\end{figure}

The afterglow of \thisgrb\ arises from the interaction of the relativistic jet with the circumburst medium, producing broadband synchrotron radiation via a decelerating external forward shock~\citep{MeszarosRees1997,Sari1998}. The observed flux follows $F_{\nu} \propto t^{-\alpha}\nu^{-\beta}$, where $\alpha$ and $\beta$ are the temporal and spectral decay indices, respectively~\citep{Sari1998}.

Optical emission from \thissn\ begins to dominate beyond ${\sim}4$~days (Figure~\ref{fig:afterglow-simple-powerlaw}), consistent with prior
results~\citep{sn2023pel, grandma}. We restrict our optical data to $t \lesssim 4$~days in all subsequent analysis. All X-ray data are used, as no supernova contribution is expected at those energies.

\begin{figure}
    \centering
    \includegraphics[width=\linewidth]{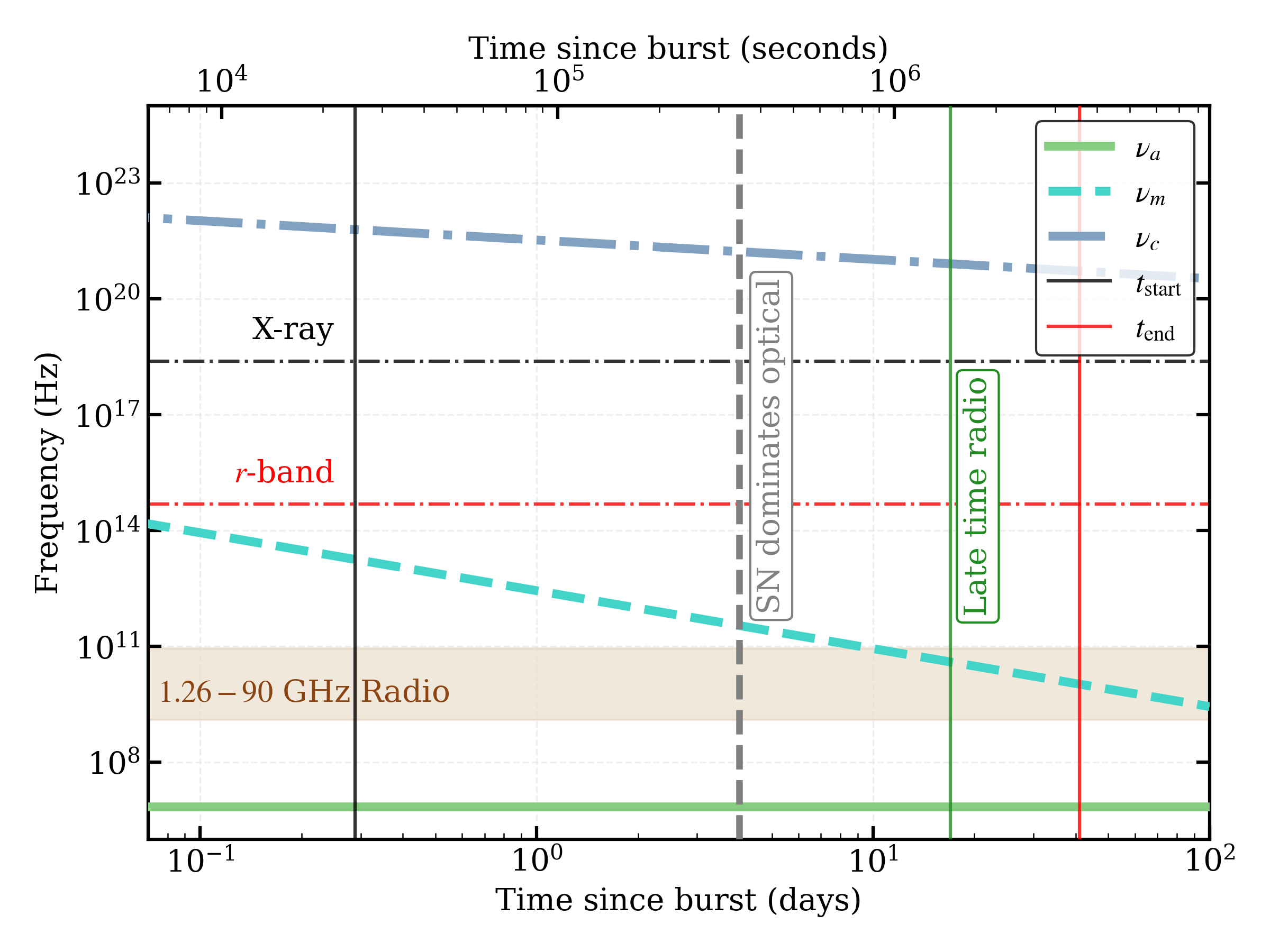}
\caption{Evolution of the synchrotron characteristic frequencies $\nu_m$, $\nu_c$, and $\nu_a$ for an on-axis jet. The black and red vertical lines mark the start and end of X-ray observations, respectively. $\nu_c$ remains above the X-ray band, consistent with the slow-cooling regime. $\nu_m$ evolves from optical to radio frequencies, crossing late-time radio observations at ~$t\ge17$~days. The self-absorption frequency $\nu_a$ remains nearly constant at $\sim$1~MHz.\label{fig:characteristic_frequencies}}
\end{figure}

The synchrotron characteristic frequencies in a constant-density (ISM) medium for an on-axis jet are
\begin{align}
\nu_m &= 3.3 \times 10^{12}
        \left(\frac{p-2}{p-1}\right)^{\!2}
        (1+z)^{1/2}\,\epsilon_B^{1/2}\,E_{52}^{1/2}\,\epsilon_e^2\,t_d^{-3/2}~\mathrm{Hz},\nonumber\\
\nu_c &= 6.3 \times 10^{15}
        (1+z)^{-1/2}\,\epsilon_B^{-3/2}\,E_{52}^{-1/2}\,n_0^{-1}\,t_d^{-1/2}~\mathrm{Hz},\nonumber\\
\nu_a &\approx 3.1 \times 10^{9}
        (1+z)^{-1}\,\epsilon_e^{-1}\,\epsilon_B^{1/5}\,E_{52}^{1/5}\,n_0^{3/5}~\mathrm{Hz},\nonumber
\end{align}
where $\nu_m$, $\nu_c$, and $\nu_a$ are the injection, cooling, and synchrotron self-absorption frequencies, respectively~\citep[Figure~\ref{fig:characteristic_frequencies};][]{Sari_1999, Sari1998}; $t_d$ is the observer time in days, $n_0$ is the ISM density in cm$^{-3}$, and $E_{52} = E_{K,\rm iso}/10^{52}$~erg. \(\epsilon_B\) is the fraction of energy in the magnetic field of the jet, \(\epsilon_e\) is the fraction of energy in the kinetic energy of the jet, and z is the redshift of the source. The values used for these parameters can be inferred from the Table~\ref{tab:afterglow_modeling}, which will be discussed in Section~\ref{sec:afterglow_modeling}. These characteristic frequencies define the synchrotron spectral regimes and their temporal evolution. After a detailed afterglow model fit is performed, we need to cross-check if the assumptions made for a given spectral band are actually consistent with the obtained results---if not, a new fit has to be carried out. Here, we present results from our best fit, where the temporal evolution of \(\nu_m\), \(\nu_c\), and \(\nu_a\) was as shown in Figure~\ref{fig:characteristic_frequencies}. Results from alternative fits are discussed in Section~\ref{sec:afterglow_alternatives}.

\subsection{X-ray Afterglow}\label{sec:xray-afterglow}
We used \swift/XRT and \chandra\ data spanning $T_0 + 7.7$~hours to $41.44$~days. The spectral model consists of a power law source and two absorption components: intrinsic source extinction and Galactic extinction, modeled with \texttt{tbabs}. The time-averaged spectrum yields a photon index $\Gamma_{\rm ph} = 1.79\pm 0.12$, corresponding to a spectral index $\beta_{\rm X} = 0.79 \pm 0.12$. No significant spectral evolution is detected in time-resolved X-ray spectral analysis. The intrinsic hydrogen column density is $N_{{\rm H},z} = (7 \pm 4)\times10^{20}~{\rm cm}^{-2}$. The X-ray temporal index is reported in Figure~\ref{fig:afterglow-simple-powerlaw}.

\subsection{Multiband Optical Afterglow}\label{sec:opt-afterglow}
We assembled optical light curves in \gband, \rband, and \iband\ from \ztf, \git, \hct, LDT, and LT (Section~\ref{sec:observations}). The optical data were corrected for Galactic extinction using the dust maps of~\citet{schlafly2011}. Host-galaxy extinction was estimated via the intrinsic neutral hydrogen column density $N_{{\rm H},z}$ from the X-ray spectral analysis (Section~\ref{sec:xray-afterglow}); using the relation $N_{{\rm H},z}/E(B-V)_z = 5.8\times10^{21}~{\rm atoms~cm^{-2}~mag^{-1}}$~\citep{cardelli1989}, we obtain an upper limit $E(B-V)_z < 0.08$~mag and neglect host extinction hereafter. The data used for modeling, along with the corrected magnitude, is given in Appendix Table~\ref{tab:afterglow_obs_table}.

Temporal decay indices from single-component power-law fits are reported in Figure~\ref{fig:afterglow-simple-powerlaw}. The optical and X-ray indices are consistent with one another, indicating no jet break within the observation window~\citep{sn2023pel, grandma}.
 
A time-averaged optical spectrum energy distribution (SED) constructed from $t \lesssim 4$~days yields $\beta_{\rm O} = 0.80 \pm 0.24$, in agreement with $\beta_{\rm X}$. The consistency of the optical and X-ray spectral indices implies that no synchrotron break frequency has crossed between the two bands during this interval, placing both in the same $\nu > \nu_c$ or $\nu_m < \nu < \nu_c$ spectral segment. Combined with the characteristic frequency evolution in Figure~\ref{fig:characteristic_frequencies}, this is consistent with the slow-cooling regime ($\nu_m < \nu_{\rm opt} < \nu_{\rm X} < \nu_c$). We analytically calculated the electron injection index \(p = 2.60 \pm 0.24\) based on the relation $\beta = (p-1)/{2}$ for the spectral index in a slow cooling regime~\citep{Sari1998}. This value is consistent with prior analyses~\citep{sn2023pel,grandma}.

\subsection{Radio Afterglow}
We included early radio data in our modeling. The early radio detections at $10$ and $15.5$~GHz ($t \lesssim 5$~days) are broadly consistent with the forward-shock model, as at those epochs $\nu_m \sim 10^{12}$~Hz (Figure~\ref{fig:characteristic_frequencies}), well above the observed radio frequencies. Both bands lie in the $\nu < \nu_m$ spectral regime where $F_\nu \propto \nu^{1/3}$, and the model reproduces the rising spectrum~\citep{2024ApJS..273...17W}. As $\nu_m \propto t^{-3/2}$ in a constant-density ISM~\citep{Sari1998}, \(\nu_m\) descends toward the GHz band over the following weeks. By $t \sim 15$--$23$~days, $\nu_m$ crosses through the $3$--$6$~GHz range, shifting the spectrum from the rising phase to the steeper post-break phase $F_\nu \propto \nu^{-(p-1)/2}$. The model, however, continues to predict rising fluxes at these frequencies, over-predicting the observed flux densities; the late-time radio data therefore fall systematically below the model curves, as seen in Figure~\ref{fig:tophat_model}. Hence, we have excluded late-time radio data obtained after $T_0$ + 17~days from our modeling.

\subsection{Afterglow Modeling}\label{sec:afterglow_modeling}
We modeled the broadband afterglow of \thisgrb\ using \texttt{jetsimpy}~\citep{2024ApJS..273...17W},
a relativistic blast-wave framework that computes synchrotron emission from a jet interacting with the circum-burst medium. \texttt{jetsimpy} approximates the blast wave as a thin 2-D shell and self-consistently accounts for lateral spreading, radiative cooling, and relativistic beaming effects. The forward shock emission is parameterized by the isotropic-equivalent kinetic energy $E_{K,\rm iso}$, the electron injection index $p$, the fraction of post-shock energy in relativistic electrons $\epsilon_e$ and magnetic field $\epsilon_B$, and the uniform ISM density $n_0$. Flux densities are evaluated with \texttt{jetsimpy}'s synchrotron radiation module (\texttt{model="sync"}) via \texttt{Jet.FluxDensity} at the luminosity distance corresponding to $z = 0.36$. Parameters were sampled in logarithmic space using \texttt{MultiNest}~\citep{2009MNRAS.398.1601F} via \texttt{PyMultiNest}~\citep{2016ascl.soft06005B} with 2000 live points, reporting the global log-evidence $\ln Z$ and posterior samples for model comparison.

\begin{figure*}[htbp!]
\centering
\begin{subfigure}[b]{\linewidth}
    \centering
    \includegraphics[width = 0.71\linewidth]{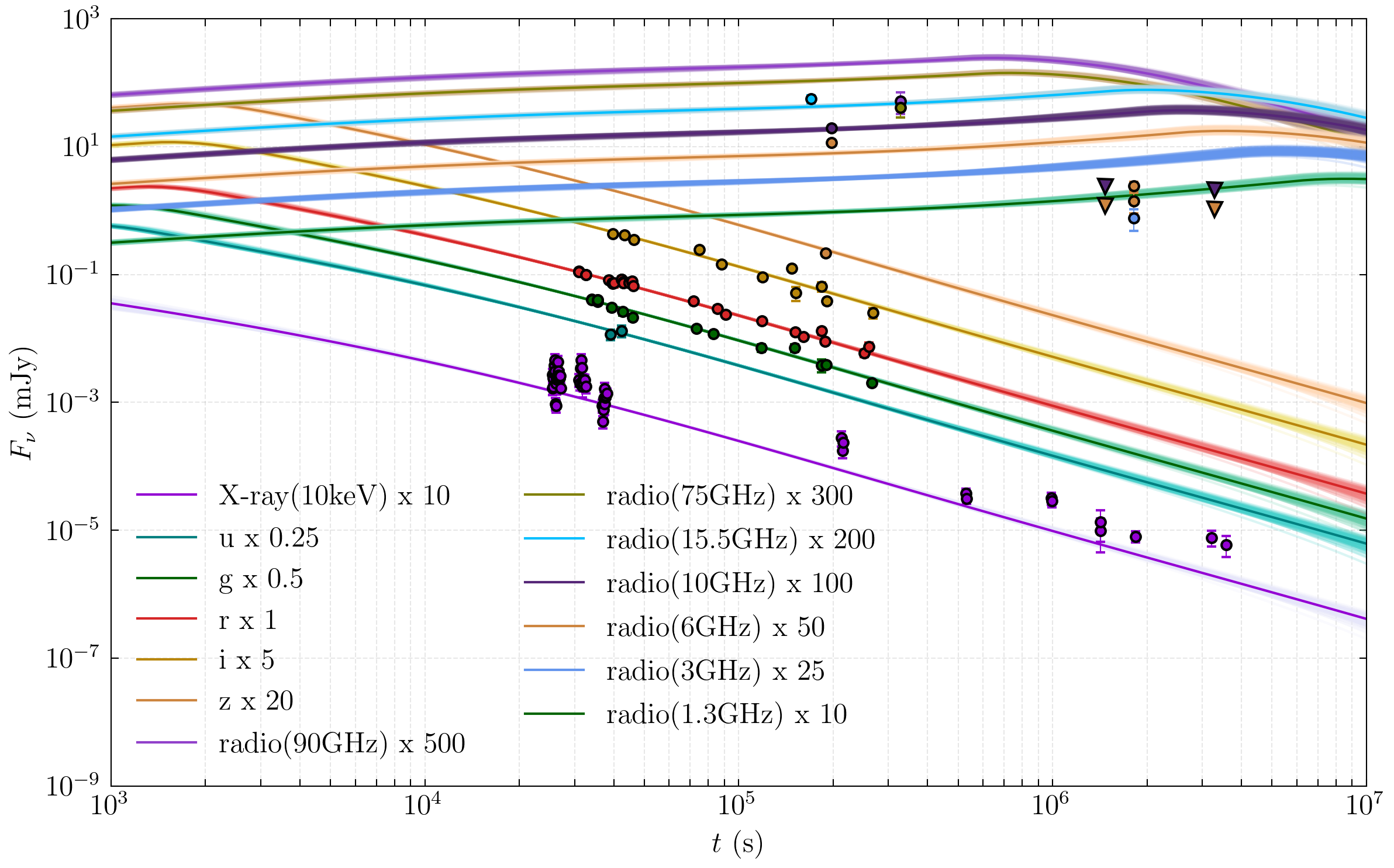}
\end{subfigure}
\begin{subfigure}[b]{\linewidth}
    \centering
    \includegraphics[width = 0.71\linewidth]{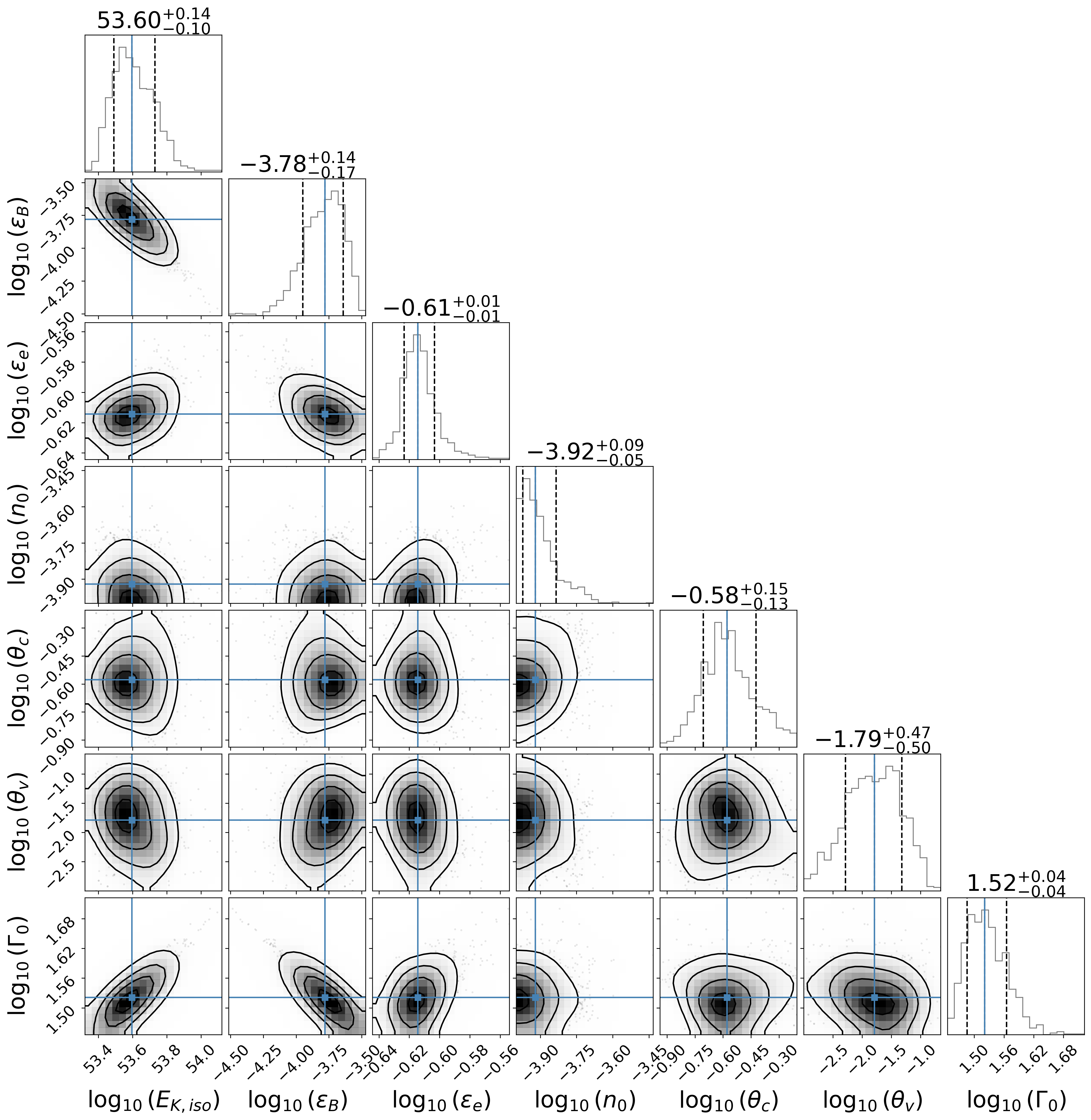}
\end{subfigure}
\caption{Model light curve and posterior constraints for \thisgrb\ using the best-fit top-hat jet model. Top: \texttt{jetsimpy} median-posterior light curves (solid lines) with
$1\sigma$ posterior bands (shaded) overlaid on multi-band observations. Downward triangles denote upper limits. Bottom: Corner plots showing the posterior distributions obtained with \textsc{MultiNest}. Blue lines indicate posterior medians.\label{fig:tophat_model}}
\end{figure*}

We explored three jet angular structures: a uniform top-hat, a power-law structured jet, and a Gaussian structured jet. For the top-hat model, the energy distribution is \(E_{\rm TopHat}(\theta) = E_{K,\rm iso}, \quad \theta \leq \theta_j,\) a uniform profile truncated sharply at the jet half-opening angle $\theta_j$ in contrast to a gradual decay for other angular structures. Further information for power-law and Gaussian jet profiles is described in \citet{pathak2026}. Models were ranked for goodness of fit using the Bayesian Information Criterion (BIC), evaluated from the nested-sampling evidence $L$, the number of free parameters $k$, and the number of flux measurements $N$. The top-hat geometry yields the lowest BIC and is adopted for all subsequent analysis; the Gaussian model is statistically indistinguishable ($\Delta\mathrm{BIC} = 1.58$), while the power-law jet is disfavored ($\Delta\mathrm{BIC} = 9.45$), as summarized in Table~\ref{tab:jet_priors_posteriors_bic}. All three geometries yield consistent parameter estimates within their respective posterior uncertainties. 

The top panel of Figure~\ref{fig:tophat_model} shows the multi-band light curves predicted by \texttt{jetsimpy} using the \textit{median} posterior parameters from \texttt{MultiNest} for this favored model, overlaid on observations; the bottom panel of Figure~\ref{fig:tophat_model} summarizes the posterior correlations. The posterior distributions are also summarized in Table~\ref{tab:afterglow_modeling}. The inferred isotropic-equivalent kinetic energy is $E_{K,\rm iso} = 4.0^{+1.5}_{-0.8} \times 10^{53}$~erg. The jet is wide, with a half-opening angle $\theta_j = 15^{+6}_{-4}$~deg, consistent with the absence of a jet break in our X-ray monitoring (Section~\ref{sec:xray-afterglow}). The line of sight is nearly on-axis at $\theta_v = 0.9^{+1.8}_{-0.6}$~deg, giving $\theta_v/\theta_j \approx 0.06$. The interstellar medium density is low, with $n_0 = 1.2^{+0.3}_{-0.1} \times 10^{-4}$~cm$^{-3}$, and the microphysical parameters are
$\epsilon_e = 0.24 \pm 0.01$ and $\epsilon_B = 1.7^{+0.6}_{-0.5} \times 10^{-4}$. The inferred initial bulk Lorentz factor is \Gzero\ $= 33 \pm 3$.
    
\begin{deluxetable}{lccc}
\tablecaption{Top-hat jet: prior ranges and MultiNest posterior constraints\label{tab:afterglow_modeling}}
\tablehead{
\colhead{Parameter} & \colhead{Prior range} &
\colhead{Posterior median} & \colhead{Units}
}
\startdata
$\log_{10} E_{K,\rm iso}$ & $[51.0,\;55.0]$ & $53.60^{+0.14}_{-0.10}$ & erg \\
$\log_{10} \epsilon_B$    & $[-5.0,\;-1.0]$ & $-3.78^{+0.14}_{-0.17}$ & \nodata \\
$\log_{10} \epsilon_e$    & $[-1.5,\;-0.5]$ & $-0.61^{+0.01}_{-0.01}$ & \nodata \\
$\log_{10} n_0$           & $[-4.0,\; 1.0]$ & $-3.92^{+0.09}_{-0.05}$ & cm$^{-3}$ \\
$\log_{10} \theta_j$      & $[-1.5,\;-0.2]$ & $-0.58^{+0.15}_{-0.13}$ & rad \\
$\log_{10} \theta_v$      & $[-3.0,\;-0.2]$ & $-1.79^{+0.47}_{-0.50}$ & rad \\
$p$                       & fixed           & $2.6$                    & \nodata \\
$\log_{10} \Gamma_0$      & $[ 1.0,\;4.0]$ & $1.52^{+0.04}_{-0.04}$  & \nodata \\
\enddata
\tablecomments{All parameters except $p$ are sampled in logarithmic space. $p = 2.6$ is fixed at the value independently inferred from the spectral index (Section~\ref{sec:opt-afterglow}). Physical equivalents of the median posteriors: $E_{K,\rm iso} = 4.0^{+1.5}_{-0.8}\times10^{53}$~erg; $\theta_j = 15^{+6}_{-4}$~deg; $\theta_v = 0.9^{+1.8}_{-0.6}$~deg; $n_0 = 1.2^{+0.3}_{-0.1}\times10^{-4}$~cm$^{-3}$; $\Gamma_0 = 33 \pm 3$.}
\end{deluxetable}
\begin{deluxetable}{lccc}
\tablecaption{Jet model comparison for three types of jet structures based on Bayesian Inference Criterion \(\mathrm{BIC} = k\ln(N) - 2L\), for nested-sampling evidence $L$, the number of free parameters $k$, and the number of flux measurements $N$. \label{tab:jet_priors_posteriors_bic}}
\tablehead{
\colhead{Model} & \colhead{BIC} & \colhead{$\Delta$BIC} & \colhead{Remark}
}
\startdata
Top-hat   & 1390.95 & 0.00 & Favored \\
Gaussian  & 1392.53 & 1.58 & Indistinguishable \\
Power law & 1400.40 & 9.45 & Disfavored \\
\enddata
\tablecomments{$\Delta$BIC is relative to the lowest-BIC model.
$\Delta\mathrm{BIC} < 2$ indicates statistically equivalent models;
$\Delta\mathrm{BIC} > 6$ indicates strong evidence against a
model~\citep{kass1995}.}
\end{deluxetable}

\subsection{Alternative Data Selection \&\ Modeling}\label{sec:afterglow_alternatives}
To assess the robustness of the top-hat fit, we performed alternative analyses varying the radio data selection and jet geometry.

Including all radio data jointly with optical and X-ray in the likelihood drives the sampler toward $E_{K,\rm iso} \approx 4 \times 10^{51}$~erg and $\epsilon_e \approx 0.49$, substantially lower energy and higher electron fraction than the best-fit values. Under these parameters, the model predicts a break in the X-ray light curve that is not observed, indicating an internal inconsistency between the radio and X-ray constraints. The inferred jet geometry is consistent with the best-fit geometry, $\theta_c \approx 11^\circ$ and $\theta_v \approx 0.4^\circ$.

Excluding radio data entirely yields the opposite bias: $E_{K,\rm iso} \approx 4 \times 10^{54}$~erg and $\epsilon_e \approx 0.05$, reflecting the well-known degeneracy between energy and electron fraction when only the optical--X-ray spectral shape constrains the microphysics. The jet geometry again remains broadly similar, with $\theta_c \approx 17^\circ$ and $\theta_v \approx 0.6^\circ$.

Replacing the top-hat profile with a Gaussian angular energy distribution yields $E_{K,\rm iso} \approx 1.6 \times 10^{54}$~erg and $\epsilon_e \approx 0.02$, with $\theta_c \approx 14^\circ$ and $\theta_v \approx 0.6^\circ$. The modeled optical light curves show no evidence of curvature attributable to off-axis structure, consistent with a nearly on-axis viewing geometry for a wide-angle jet.

Across all three alternatives and the best-fit, the inferred jet geometry is robust: a wide-angle jet
($\theta_c \approx 11^\circ$--$17^\circ$) viewed close to the jet axis ($\theta_v \lesssim 1^\circ$), with no jet break within the monitored window. The primary sensitivity is in the parameters $E_{K,\rm iso}$ and $\epsilon_e$, which are dependent on the radio data selection. The fiducial fit using early radio data alongside optical data (Section~\ref{sec:afterglow_modeling}) yields a solution that is simultaneously consistent with the optical, X-ray, and early radio observations.

\section{Discussion}\label{sec:discussion}

\subsection{Absence of Jet Break}
The extensive afterglow follow-up of \thisgrb\ in X-ray and optical shows a long-lived afterglow, best fit by a simple power law decay with no further steepening at late times (Figure~\ref{fig:afterglow-simple-powerlaw}). Two phenomenological scenarios could explain this: (1) An early jet break occurring \(\approx 7\)~hours after the burst start time, or (2) a late jet break occurring after the last observation at \(\approx 41\)~days since burst. Our broadband afterglow fits yield an X‑ray temporal decay index $\alpha_\mathrm{X} = 1.23 \pm 0.04$ and an X‑ray spectral index $\beta_\mathrm{X} = 0.79 \pm 0.12$, while the optical decay indices are $\alpha_\mathrm{r} = 1.28 \pm 0.01$ (\rband\ filter) and $\alpha_\mathrm{g} = 1.31 \pm 0.01$ (\gband\ filter), with an optical spectral index $\beta_\mathrm{O} = 0.80 \pm 0.24$. The inferred electron injection index is $p = 2.6 \pm 0.24$. Using the standard on‑axis jet break formalism~\citep{Sari_1999}, we find that $\theta_j \leq 1.1^\circ$ if the jet break occurred before 7~hours, or $\theta_j \gtrsim 7.2^\circ$ if the jet break happened after 41~days. These two cases correspond to either a narrow, highly collimated jet viewed close to the jet axis or a wide jet with a small viewing angle.

The first scenario, in which the jet break occurs before our earliest observation at ${\sim}7$~hours, would require a post-break temporal decay $\alpha = -p$~\citep{Sari_1999}, implying $p \approx 1.2$ from the observed slope. This is inconsistent with the value $p \approx 2.6$ independently inferred from the spectral index $\beta = (p-1)/2$ (Section~\ref{sec:afterglow-analysis}). The two estimates of $p$ differ by more than a factor of two, violating the closure relations~\citep{racusin2009}, and this scenario is therefore ruled out. In contrast, the second scenario where jet break occurs post \(\sim 41\)~days would require a wide jet opening angle, which matches well with the inferred value of jet half-opening angle of \(\theta_j \approx 15\)~degrees and viewing angle of \(\theta_v \approx 1\)~degrees. The isotropic-equivalent kinetic energy of the jet is \ekiso\(= 4.0_{-0.8}^{+1.5} \times 10^{53}\)~erg which after correcting for beaming, gives \(E_k \approx 1.4 \times 10^{52}\)~erg.


\subsection{High Energy Polarization in Perspective}\label{sec:pol_discussion}


The physical interpretation of prompt polarization fundamentally depends on the geometry of the visible emitting region, defined primarily by the relativistic beaming angle ($\Gzero^{-1}$), the observer's viewing angle (\thetav), and the jet half-opening angle (\thetaj). Our multiwavelength afterglow modeling indicates the jet half-opening angle \(\theta_j = 15^{+6}_{-4}\)~deg, the viewing angle \(\theta_v = 0.9^{+1.8}_{-0.6}\)~deg, and the initial bulk Lorentz factor \Gzero\(=33 \pm 3\) giving \(1/\Gamma_0 \ll \theta_j\). Constraints of \Gzero\ = 315~\citep{ror2024promptafterglowanalysisfermilat}, 70~\citep{mohnani2026broad} are also consistent with an ultrarelativistic interpretation with the observer well within the jet cone angle \(\theta_v \ll \theta_j\) for a top-hat jet. The polarization predictions for this geometric regime are summarized in Figure~1 of \citet{gill2021}, and are discussed below. 


Photospheric emission (P) reaches at most $\Pi \approx 40\%$ for the most favorable cases of narrow jets with steep Lorentz factor gradients and $\theta_v \gtrsim \theta_j$~\citep{lundman2014}; for $\theta_v \ll \theta_j$ as in our case, the azimuthal symmetry of the photosphere is nearly exact and the expected $\Pi$ is a few percent~\citep{lundman2014,toma2009}. Synchrotron with a random magnetic field (SR; $B_\perp$) produces $4\% \lesssim \Pi \lesssim 28\%$ only for lines of sight grazing the jet edge ($|\theta_v - \theta_j| \sim 1/\Gamma$)~\citep{gill2021}; for $\theta_v \ll \theta_j$ azimuthal cancellation drives $\Pi \to 0$. In contrast, the synchrotron with a globally ordered toroidal magnetic field (SO; $B_{\rm tor}$) naturally breaks the azimuthal symmetry for any $1/\Gamma \ll \theta_j$, and a measurement $\Pi \gtrsim 50\%$ signals the presence of such an ordered field~\citep{gill2021,toma2009}. Compton drag (CD) offers a comparable signature, also predicting $\Pi \sim 10-50\%$ for viewing angles comparable to the jet half-opening angle, but leads $\Pi \to 0$ for $\theta_v \ll \theta_j$~\citep{toma2009,gill2021}.
 
\astrosat/CZTI shows hints of high polarization, with \(\Pi \ge 50\%\) at 1\(\sigma\) for a Bayes factor of \(\approx 2\) in energy range \(300\)--\(600\)~keV. Because of detector saturation during the burst peak, this analysis was restricted to time-integrated data from the post-peak interval, yielding a marginal polarization signature. This lower-limit measurement, along with jet-geometry hints, suggests incompatibility with the near-zero \pf\ prediction for SR, P, and CD. This result favors an ordered magnetic field with synchrotron emission (SO), hinting at a jet with coherent field structure on angular scales $\gtrsim 1/\Gamma$ around the line of sight. The upper limit of $\Pi \lesssim 64\%$ at $99\%$ in the energy range $100$--$600$~keV remains consistent with the \(300\)--\(600\)~keV lower limit.

This highlights an important, often overlooked design requirement for future polarimeters: the priority is not only to be more sensitive to faint bursts but also to have sufficient dynamic range to avoid saturation by the brightest bursts like \thisgrb\ or GRB\,230307A. Bright, well-localized GRBs like \thisgrb\ offer a chance to undertake detailed spectro-polarimetric studies but are most vulnerable to saturation losses. Upcoming missions like \emph{Daksha}~\citep{bhalerao2022daksha}, \emph{COSI}~\citep{tomsick2022cosi}, and \emph{POLAR-2}~\citep{kole2020polar2}, designed with improved dynamic range and wide-field coverage, are better positioned to exploit such exceptional events.

\begin{acknowledgments}
CZT--Imager is the result of a collaborative effort involving multiple institutes across India. The Tata Institute of Fundamental Research in Mumbai played a central role in spearheading the instrument's design and development. The Vikram Sarabhai Space Centre in Thiruvananthapuram contributed to electronic design, assembly, and testing, while the ISRO Satellite Centre (ISAC) in Bengaluru provided expertise in mechanical design, quality consultation, and project management.
The Inter University Centre for Astronomy and Astrophysics (IUCAA) in Pune was responsible for the Coded Mask design, instrument calibration, and the operation of the Payload Operation Centre. The Space Application Centre (SAC) in Ahmedabad supplied the essential analysis software, and the Physical Research Laboratory (PRL) in Ahmedabad contributed the polarization detection algorithm and conducted ground calibration. Several industries were actively involved in the fabrication process, and the university sector played a crucial role in testing and evaluating the payload.
The Indian Space Research Organisation (ISRO) not only funded the project but also provided essential management and facilitation throughout its development.

We thank all members of the GROWTH collaboration for helping with observations and data processing.

The GROWTH India Telescope \citep[GIT; ][]{2022AJ....164...90K} is a 70-cm telescope with a 0.7-degree field of view, set up by the Indian Institute of Astrophysics (IIA) and the Indian Institute of Technology Bombay (IITB) with funding from DST-SERB and IUSSTF. It is located at the Indian Astronomical Observatory (Hanle), operated by IIA. We acknowledge funding by the IITB alumni batch of 1994, which partially supports the operations of the telescope. Telescope technical details are available at \url{https://sites.google.com/view/growthindia/}.

This work is partially based on data obtained with the 2m Himalayan Chandra Telescope of the Indian Astronomical Observatory (IAO). We thank the staff of IAO, Hanle, and CREST, Hosakote, that made these observations possible. The facilities at IAO and CREST are operated by the Indian Institute of Astrophysics, Bangalore.

M.W.C. acknowledges support from the National Science Foundation with grant numbers PHY-2117997, PHY-2308862 and PHY-2409481.
I.A. is supported by the National Science Foundation award AST 2505775, NASA grant 24-ADAP24- 0159, Scialog award SA-LSST-2024-102a and LSST2025-112b.
G.C. Anupama acknowledges support from the Indian National Science Academy (INSA) under its Senior Scientist programme.
O.J.R. gratefully acknowledges NASA funding through contract 80MSFC17M0022 and Research Ireland Pathway Funding through contract 24/PATH-S/12742(T).

Based on observations obtained with the Samuel Oschin Telescope 48-inch and the 60-inch Telescope at the Palomar Observatory as part of the Zwicky Transient Facility project. ZTF is supported by the National Science Foundation under Grant No. AST-2034437 and a collaboration including Caltech, IPAC, the Weizmann Institute of Science, the Oskar Klein Center at Stockholm University, the University of Maryland, Deutsches Elektronen-Synchrotron and Humboldt University, the TANGO Consortium of Taiwan, the University of Wisconsin at Milwaukee, Trinity College Dublin, Lawrence Livermore National Laboratories, IN2P3, University of Warwick, Ruhr University Bochum, Cornell University, and Northwestern University. Operations are conducted by COO, IPAC, and UW.

SED Machine is based upon work supported by the National Science Foundation under Grant No. 1106171. We thank Anna Ho for triggering P60 for observing this GRB.

This work made use of data supplied by the UK Swift Science Data Centre at the University of Leicester.
The IIT Bombay team sincerely thanks BDP UGL Global Logistics Pvt. Ltd. for their generous CSR support toward our computational needs.
Special thanks to Tanishk Mohan for collating all afterglow data and Harsh Kumar for working on the Chandra proposal. 

We thank the CXO staff---in particular, Patrick Slane, Dan Schwartz, Harvey Tananbaum, Steiner James, Doug Swartz, and Malgorzata Sobolewska for rapidly approving and planning this observation.
\end{acknowledgments}

\begin{contribution}

All authors contributed equally to this collaborative work.


\end{contribution}

%
\facilities{\fermi\ (GBM), \astrosat\ (CZTI), \swift\ (XRT and UVOT), \chandra, GIT:0.7\,m, HCT:2\,m, ZTF, SEDM.}

\software{Astropy~\citep{2013A&A...558A..33A,2018AJ....156..123A,2022ApJ...935..167A}, Source Extractor~\citep{1996A&AS..117..393B}, Astro-SCRAPPY~\citep{2019ascl.soft07032M}, \sw{solve-field} astrometry engine~\citep{2010AJ....139.1782L}, \sw{PSFEx}~\citep{2013ascl.soft01001B},  \jetsimpy~\citep{2024ApJS..273...17W}, \sw{PyMultiNest}~\citep{2014A&A...564A.125B}, \sw{3ML}~\citep{2015arXiv150708343V}, \texttt{dustmaps}~\citep{2018JOSS....3..695M}.}



\appendix
\startlongtable
\begin{deluxetable*}{ccccccc}
\tablecaption{Multi-wavelength afterglow observations of \thisgrb\ across the optical, X-ray, and radio bands. The table presents the time elapsed since the burst ($T-T_0$) in seconds, filter/band information, central frequency (Hz), unabsorbed flux density (mJy), Galactic extinction-corrected magnitude (AB system)~\cite{schlafly2011}, observing instrument, and the corresponding references for each data point.\label{tab:afterglow_obs_table}} 
\tabletypesize{\scriptsize}
\tablehead{
\colhead{Time $-$ $T_0$} & \colhead{Filter} & \colhead{Frequency} & \colhead{Flux} & \colhead{Corr Mag} & \colhead{Instrument} & \colhead{Ref.} \\
\colhead{(sec)} & \colhead{} & \colhead{($\times 10^{14}$ Hz)} & \colhead{mJy} & \colhead{AB} & \colhead{} & \colhead{}
}
\startdata
25464.91 & 10~keV & 24180.0 & $(1.66 \pm 0.37) \times 10^{-4}$ & - & \swift/XRT & \cite{gcn34400} \\
25542.63 & 10~keV & 24180.0 & $(2.74 \pm 0.60) \times 10^{-4}$ & - & \swift/XRT & \cite{gcn34400} \\
25601.81 & 10~keV & 24180.0 & $(2.37 \pm 0.53) \times 10^{-4}$ & - & \swift/XRT & \cite{gcn34400} \\
25676.58 & 10~keV & 24180.0 & $(1.73 \pm 0.38) \times 10^{-4}$ & - & \swift/XRT & \cite{gcn34400} \\
25768.42 & 10~keV & 24180.0 & $(2.92 \pm 0.66) \times 10^{-4}$ & - & \swift/XRT & \cite{gcn34400} \\
25837.50 & 10~keV & 24180.0 & $(3.81 \pm 0.86) \times 10^{-4}$ & - & \swift/XRT & \cite{gcn34400} \\
25919.32 & 10~keV & 24180.0 & $(4.61 \pm 1.01) \times 10^{-4}$ & - & \swift/XRT & \cite{gcn34400} \\
25986.69 & 10~keV & 24180.0 & $(3.06 \pm 0.69) \times 10^{-4}$ & - & \swift/XRT & \cite{gcn34400} \\
26061.87 & 10~keV & 24180.0 & $(9.57 \pm 2.09) \times 10^{-5}$ & - & \swift/XRT & \cite{gcn34400} \\
26167.21 & 10~keV & 24180.0 & $(8.93 \pm 2.01) \times 10^{-5}$ & - & \swift/XRT & \cite{gcn34400} \\
26239.05 & 10~keV & 24180.0 & $(1.99 \pm 0.45) \times 10^{-4}$ & - & \swift/XRT & \cite{gcn34400} \\
26265.60 & 10~keV & 24180.0 & $(2.29 \pm 0.95) \times 10^{-4}$ & - & \swift/XRT & \cite{gcn34400} \\
26315.40 & 10~keV & 24180.0 & $(2.48 \pm 0.56) \times 10^{-4}$ & - & \swift/XRT & \cite{gcn34400} \\
26410.54 & 10~keV & 24180.0 & $(2.58 \pm 0.58) \times 10^{-4}$ & - & \swift/XRT & \cite{gcn34400} \\
26493.54 & 10~keV & 24180.0 & $(4.34 \pm 0.97) \times 10^{-4}$ & - & \swift/XRT & \cite{gcn34400} \\
26586.30 & 10~keV & 24180.0 & $(3.11 \pm 0.70) \times 10^{-4}$ & - & \swift/XRT & \cite{gcn34400} \\
26676.31 & 10~keV & 24180.0 & $(2.65 \pm 0.60) \times 10^{-4}$ & - & \swift/XRT & \cite{gcn34400} \\
26808.70 & 10~keV & 24180.0 & $(2.42 \pm 0.55) \times 10^{-4}$ & - & \swift/XRT & \cite{gcn34400} \\
26917.35 & 10~keV & 24180.0 & $(2.47 \pm 0.54) \times 10^{-4}$ & - & \swift/XRT & \cite{gcn34400} \\
27009.43 & 10~keV & 24180.0 & $(2.63 \pm 0.59) \times 10^{-4}$ & - & \swift/XRT & \cite{gcn34400} \\
27118.05 & 10~keV & 24180.0 & $(1.69 \pm 0.31) \times 10^{-4}$ & - & \swift/XRT & \cite{gcn34400} \\
31017.60 & \rband\ & 4.8384 & 0.110$\pm$0.004 & 18.80$\pm$0.04 & ZTF & \cite{sn2023pel} \\
31017.60 & \rband\ & 4.8384 & 0.112$\pm$0.002 & 18.78$\pm$0.02 & ZTF & \cite{sn2023pel} \\
31017.60 & \rband\ & 4.8384 & 0.110$\pm$0.005 & 18.80$\pm$0.05 & ZTF & \cite{grandma} \\
31016.76 & 10~keV & 24180.0 & $(2.25 \pm 0.51) \times 10^{-4}$ & - & \swift/XRT & \cite{gcn34400} \\
31182.98 & 10~keV & 24180.0 & $(1.98 \pm 0.44) \times 10^{-4}$ & - & \swift/XRT & \cite{gcn34400} \\
31316.51 & 10~keV & 24180.0 & $(3.46 \pm 0.78) \times 10^{-4}$ & - & \swift/XRT & \cite{gcn34400} \\
31479.11 & 10~keV & 24180.0 & $(4.59 \pm 1.03) \times 10^{-4}$ & - & \swift/XRT & \cite{gcn34400} \\
31623.79 & 10~keV & 24180.0 & $(3.48 \pm 0.78) \times 10^{-4}$ & - & \swift/XRT & \cite{gcn34400} \\
31780.24 & 10~keV & 24180.0 & $(1.93 \pm 0.43) \times 10^{-4}$ & - & \swift/XRT & \cite{gcn34400} \\
31795.20 & 10~keV & 24180.0 & $(1.77 \pm 0.57) \times 10^{-4}$ & - & \swift/XRT & \cite{gcn34400} \\
31934.82 & 10~keV & 24180.0 & $(2.27 \pm 0.51) \times 10^{-4}$ & - & \swift/XRT & \cite{gcn34400} \\
32077.99 & 10~keV & 24180.0 & $(1.92 \pm 0.43) \times 10^{-4}$ & - & \swift/XRT & \cite{gcn34400} \\
32248.88 & 10~keV & 24180.0 & $(2.23 \pm 0.50) \times 10^{-4}$ & - & \swift/XRT & \cite{gcn34400} \\
32572.80 & \rband\ & 4.8384 & 0.100$\pm$0.003 & 18.90$\pm$0.03 & ZTF & \cite{sn2023pel} \\
32572.80 & \rband\ & 4.8384 & 0.099$\pm$0.001 & 18.91$\pm$0.01 & ZTF & \cite{sn2023pel} \\
32551.62 & 10~keV & 24180.0 & $(1.79 \pm 0.39) \times 10^{-4}$ & - & \swift/XRT & \cite{gcn34400} \\
33955.20 & \gband\ & 6.249 & 0.082$\pm$0.002 & 19.11$\pm$0.03 & ZTF & \cite{sn2023pel} \\
33955.20 & \gband\ & 6.249 & 0.081$\pm$0.001 & 19.13$\pm$0.01 & ZTF & \cite{sn2023pel} \\
35510.40 & \gband\ & 6.249 & 0.080$\pm$0.002 & 19.14$\pm$0.03 & ZTF & \cite{sn2023pel} \\
35510.40 & \gband\ & 6.249 & 0.075$\pm$0.002 & 19.21$\pm$0.03 & ZTF & \cite{sn2023pel} \\
35510.40 & \gband\ & 6.249 & 0.080$\pm$0.007 & 19.14$\pm$0.10 & ZTF & \cite{grandma} \\
36729.28 & 10~keV & 24180.0 & $(8.90 \pm 1.99) \times 10^{-5}$ & - & \swift/XRT & \cite{gcn34400} \\
36850.48 & 10~keV & 24180.0 & $(5.00 \pm 1.12) \times 10^{-5}$ & - & \swift/XRT & \cite{gcn34400} \\
37037.54 & 10~keV & 24180.0 & $(7.73 \pm 1.74) \times 10^{-5}$ & - & \swift/XRT & \cite{gcn34400} \\
37144.62 & 10~keV & 24180.0 & $(1.18 \pm 0.26) \times 10^{-4}$ & - & \swift/XRT & \cite{gcn34400} \\
37295.04 & 10~keV & 24180.0 & $(9.65 \pm 2.17) \times 10^{-5}$ & - & \swift/XRT & \cite{gcn34400} \\
37324.80 & 10~keV & 24180.0 & $(9.55 \pm 3.25) \times 10^{-5}$ & - & \swift/XRT & \cite{gcn34400} \\
37441.36 & 10~keV & 24180.0 & $(1.65 \pm 0.36) \times 10^{-4}$ & - & \swift/XRT & \cite{gcn34400} \\
37581.88 & 10~keV & 24180.0 & $(1.19 \pm 0.27) \times 10^{-4}$ & - & \swift/XRT & \cite{gcn34400} \\
37765.90 & 10~keV & 24180.0 & $(1.28 \pm 0.29) \times 10^{-4}$ & - & \swift/XRT & \cite{gcn34400} \\
38018.37 & 10~keV & 24180.0 & $(1.39 \pm 0.23) \times 10^{-4}$ & - & \swift/XRT & \cite{gcn34400} \\
38534.40 & \rband\ & 4.8384 & 0.083$\pm$0.002 & 19.11$\pm$0.02 & ZTF & \cite{sn2023pel} \\
38966.40 & \uband\ & 8.1178 & 0.046$\pm$0.008 & 19.74$\pm$0.20 & SEDM & \cite{sn2023pel} \\
39312.00 & \gband\ & 6.249 & 0.061$\pm$0.007 & 19.44$\pm$0.13 & SEDM & \cite{sn2023pel} \\
39484.80 & \rband\ & 4.8384 & 0.074$\pm$0.003 & 19.23$\pm$0.04 & SEDM & \cite{sn2023pel} \\
39744.00 & \rband\ & 4.8384 & 0.073$\pm$0.006 & 19.25$\pm$0.09 & ZTF & \cite{sn2023pel} \\
39744.00 & \iband\ & 3.989 & 0.086$\pm$0.004 & 19.06$\pm$0.05 & SEDM & \cite{sn2023pel} \\
40003.20 & \rband\ & 4.8384 & 0.074$\pm$0.011 & 19.23$\pm$0.16 & SEDM & \cite{sn2023pel} \\
42249.60 & \rband\ & 4.8384 & 0.084$\pm$0.009 & 19.09$\pm$0.12 & ZTF & \cite{sn2023pel} \\
42249.60 & \rband\ & 4.8384 & 0.083$\pm$0.005 & 19.10$\pm$0.06 & ZTF & \cite{sn2023pel} \\
42336.00 & \uband\ & 8.1178 & 0.053$\pm$0.011 & 19.59$\pm$0.23 & SEDM & \cite{sn2023pel} \\
42681.60 & \gband\ & 6.249 & 0.052$\pm$0.007 & 19.60$\pm$0.14 & SEDM & \cite{sn2023pel} \\
42940.80 & \rband\ & 4.8384 & 0.073$\pm$0.003 & 19.24$\pm$0.04 & SEDM & \cite{sn2023pel} \\
42940.80 & \rband\ & 4.8384 & 0.077$\pm$0.011 & 19.19$\pm$0.16 & ZTF & \cite{sn2023pel} \\
42940.80 & \rband\ & 4.8384 & 0.075$\pm$0.005 & 19.22$\pm$0.07 & ZTF & \cite{sn2023pel} \\
43200.00 & \iband\ & 3.989 & 0.082$\pm$0.004 & 19.11$\pm$0.05 & SEDM & \cite{sn2023pel} \\
44755.20 & \rband\ & 4.8384 & 0.074$\pm$0.007 & 19.23$\pm$0.10 & ZTF & \cite{sn2023pel} \\
44755.20 & \rband\ & 4.8384 & 0.075$\pm$0.005 & 19.21$\pm$0.07 & ZTF & \cite{sn2023pel} \\
45792.00 & \rband\ & 4.8384 & 0.075$\pm$0.008 & 19.21$\pm$0.12 & ZTF & \cite{sn2023pel} \\
45792.00 & \rband\ & 4.8384 & 0.079$\pm$0.004 & 19.16$\pm$0.06 & ZTF & \cite{sn2023pel} \\
45878.40 & \gband\ & 6.249 & 0.043$\pm$0.005 & 19.81$\pm$0.13 & SEDM & \cite{sn2023pel} \\
46137.60 & \rband\ & 4.8384 & 0.066$\pm$0.002 & 19.35$\pm$0.03 & SEDM & \cite{sn2023pel} \\
46310.40 & \iband\ & 3.989 & 0.070$\pm$0.003 & 19.28$\pm$0.04 & SEDM & \cite{sn2023pel} \\
71884.80 & \rband\ & 4.8384 & 0.039$\pm$0.002 & 19.93$\pm$0.05 & GIT & \cite{sn2023pel} \\
73440.00 & \gband\ & 6.249 & 0.028$\pm$0.002 & 20.26$\pm$0.06 & GIT & \cite{sn2023pel} \\
74995.20 & \iband\ & 3.989 & 0.049$\pm$0.003 & 19.67$\pm$0.06 & GIT & \cite{sn2023pel} \\
82944.00 & \gband\ & 6.249 & 0.024$\pm$0.002 & 20.47$\pm$0.08 & KAO & \cite{grandma} \\
85449.60 & \rband\ & 4.8384 & 0.029$\pm$0.003 & 20.24$\pm$0.12 & KAO & \cite{grandma} \\
88214.40 & \iband\ & 3.989 & 0.029$\pm$0.002 & 20.25$\pm$0.06 & KAO & \cite{grandma} \\
90720.00 & \rband\ & 4.8384 & 0.024$\pm$0.001 & 20.46$\pm$0.06 & T193/MISTRAL & \cite{grandma} \\
118195.20 & \gband\ & 6.249 & 0.014$\pm$0.002 & 21.01$\pm$0.15 & SEDM & \cite{sn2023pel} \\
118540.80 & \rband\ & 4.8384 & 0.019$\pm$0.001 & 20.71$\pm$0.07 & SEDM & \cite{sn2023pel} \\
118886.40 & \iband\ & 3.989 & 0.018$\pm$0.002 & 20.75$\pm$0.12 & SEDM & \cite{sn2023pel} \\
147312.00 & \iband\ & 3.989 & 0.025$\pm$0.001 & 20.40$\pm$0.04 & CAHA & \cite{grandma} \\
150595.20 & \gband\ & 6.249 & 0.014$\pm$0.002 & 21.01$\pm$0.15 & GMG & \cite{grandma} \\
151200.00 & \rband & 4.8384 & 0.013$\pm$0.001 & 21.15$\pm$0.08 & GMG & \cite{grandma} \\
151891.20 & \iband\ & 3.989 & 0.010$\pm$0.003 & 21.37$\pm$0.27 & GMG & \cite{grandma} \\
160704.00 & \rband\ & 4.8384 & 0.011$\pm$0.001 & 21.33$\pm$0.05 & GIT & \cite{sn2023pel} \\
170139.00 & 15.5~GHz & 0.000155 & 0.28$\pm$0.04 & 17.78$\pm$0.16 & AMI-LA & \cite{gcn34433} \\
183600.00 & \gband\ & 6.249 & 0.008$\pm$0.002 & 21.69$\pm$0.24 & LT & \cite{sn2023pel} \\
183772.80 & \rband\ & 4.8384 & 0.013$\pm$0.002 & 21.11$\pm$0.14 & LT & \cite{sn2023pel} \\
183945.60 & \iband\ & 3.989 & 0.013$\pm$0.002 & 21.11$\pm$0.18 & LT & \cite{sn2023pel} \\
188784.00 & \rband\ & 4.8384 & 0.0090$\pm$0.0005 & 21.52$\pm$0.06 & NOT & \cite{grandma} \\
189129.60 & \zband\ & 3.4138 & 0.011$\pm$0.001 & 21.31$\pm$0.12 & NOT & \cite{grandma} \\
189820.80 & \gband\ & 6.249 & 0.0077$\pm$0.0003 & 21.68$\pm$0.04 & NOT & \cite{grandma} \\
190512.00 & \iband\ & 3.989 & 0.0077$\pm$0.0008 & 21.69$\pm$0.11 & NOT & \cite{grandma} \\
197652.00 & 6~GHz & 0.00006 & 0.230$\pm$0.010 & 18.00$\pm$0.05 & VLA & \cite{gcn34552} \\
197652.00 & 10~GHz & 0.0001 & 0.196$\pm$0.007 & 18.17$\pm$0.04 & VLA & \cite{gcn34552} \\
212307.50 & 10~keV & 24180.0 & $(2.77 \pm 0.73) \times 10^{-5}$ & - & \swift/XRT & \cite{gcn34400} \\
214012.80 & 10~keV & 24180.0 & $(1.77 \pm 0.42) \times 10^{-5}$ & - & \swift/XRT & \cite{gcn34400} \\
214919.50 & 10~keV & 24180.0 & $(2.34 \pm 0.47) \times 10^{-5}$ & - & \swift/XRT & \cite{gcn34400} \\
250646.40 & \rband\ & 4.8384 & 0.0059$\pm$0.0008 & 21.97$\pm$0.15 & NOWT & \cite{grandma} \\
259977.60 & \rband\ & 4.8384 & 0.0075$\pm$0.0011 & 21.72$\pm$0.16 & HCT & \cite{sn2023pel} \\
265852.80 & \gband\ & 6.249 & 0.0040$\pm$0.0004 & 22.39$\pm$0.10 & CAHA & \cite{grandma} \\
268358.40 & \iband\ & 3.989 & 0.0051$\pm$0.0010 & 22.14$\pm$0.21 & CAHA & \cite{grandma} \\
327048.00 & 75~GHz & 0.00075 & 0.138$\pm$0.042 & 18.55$\pm$0.33 & NOEMA & \cite{gcn34468} \\
327048.00 & 90~GHz & 0.0009 & 0.103$\pm$0.038 & 18.87$\pm$0.40 & NOEMA & \cite{gcn34468} \\
530259.90 & 10~keV & 24180.0 & $(3.78 \pm 0.69) \times 10^{-6}$ & - & \swift/XRT & \cite{gcn34400} \\
531792.00 & 10~keV & 24180.0 & $(3.16 \pm 0.58) \times 10^{-6}$ & - & \swift/XRT & \cite{gcn34400} \\
990768.50 & 10~keV & 24180.0 & $(3.18 \pm 0.68) \times 10^{-6}$ & - & \swift/XRT & \cite{gcn34400} \\
993859.20 & 10~keV & 24180.0 & $(2.88 \pm 0.61) \times 10^{-6}$ & - & \swift/XRT & \cite{gcn34400} \\
1421366.0 & 10~keV & 24180.0 & $(9.91 \pm 5.38) \times 10^{-7}$ & - & \swift/XRT & \cite{gcn34400} \\
1421373.0 & 10~keV & 24180.0 & $(1.35 \pm 0.68) \times 10^{-6}$ & - & \swift/XRT & \cite{gcn34400} \\
1468800.0 & 6~GHz & 0.00006 & $<0.024$ & - & VLA & \cite{gcn35505} \\
1468800.0 & 10~GHz & 0.0001 & $<0.024$ & - & VLA & \cite{gcn35505} \\
1812399.0 & 3~GHz & 0.00003 & 0.031$\pm$0.011 & 20.19$\pm$0.40 & VLA & \cite{gcn34735} \\
1812399.0 & 6~GHz & 0.00006 & 0.049$\pm$0.008 & 19.67$\pm$0.17 & VLA & \cite{gcn34735} \\
1812399.0 & 6~GHz & 0.00006 & 0.028$\pm$0.010 & 20.27$\pm$0.39 & VLA & \cite{gcn34735} \\
1836371.0 & 10~keV & 24180.0 & $(7.98 \pm 1.58) \times 10^{-7}$ & - & \chandra & \cite{gcn34632} \\
3083508.0 & 1.26~GHz & 0.0000126 & $<0.04$ & $>$19.82 & uGMRT & \cite{gcn34735, gcn34727} \\
3206883.0 & 10~keV & 24180.0 & $(7.70 \pm 2.14) \times 10^{-7}$ & - & \chandra & This paper \\
3283200.0 & 6~GHz & 0.00006 & $<0.021$ & - & VLA & \cite{gcn35505} \\
3283200.0 & 10~GHz & 0.0001 & $<0.021$ & - & VLA & \cite{gcn35505} \\
3570113.0 & 10~keV & 24180.0 & $(5.97 \pm 2.18) \times 10^{-7}$ & - & \chandra & This paper \\
\enddata
\end{deluxetable*}


\bibliography{sample701}{}
\bibliographystyle{aasjournalv7}



\end{document}